\documentclass[a4paper,twocolumn,superscriptaddress,pre, aps, floatfix ]{revtex4-1}
\usepackage{amssymb}
\usepackage{amsmath}
\usepackage{marvosym}
\usepackage{graphicx}
\usepackage{url}

\begin{document}
\title{Memory-induced complex contagion in epidemic spreading}
\author{Xavier R. Hoffmann}
\email{xrhoffmann@gmail.com}
\affiliation{Departament de F\'isica de la Mat\`eria Condensada, Universitat de Barcelona, Spain}
\affiliation{Universitat de Barcelona Institute of Complex Systems (UBICS), Spain}
\author{Mari\'an Bogu\~n\'a}
\email{marian.boguna@ub.edu}
\affiliation{Departament de F\'isica de la Mat\`eria Condensada, Universitat de Barcelona, Spain}
\affiliation{Universitat de Barcelona Institute of Complex Systems (UBICS), Spain}

\begin{abstract}
Albeit epidemic models have evolved into powerful predictive tools for the spread of diseases and opinions, most assume memoryless agents and independent transmission channels. We develop an infection mechanism that is endowed with memory of past exposures and simultaneously incorporates the joint effect of multiple infectious sources. Analytic equations and simulations of the susceptible-infected-susceptible model in unstructured substrates reveal the emergence of an additional phase that separates the usual healthy and endemic ones. This intermediate phase shows fundamentally distinct characteristics, and the system exhibits either excitability or an exotic variant of bistability. Moreover, the transition to endemicity presents hybrid aspects. These features are the product of an intricate balance between two memory modes and indicate that non-Markovian effects significantly alter the properties of  spreading processes.
\end{abstract}

\maketitle
\section{Introduction}
Epidemic modeling has proven to be a powerful tool for the study of contagion phenomena in biological, social, and technological systems, e.g., diseases, opinions, rumors and innovation~\cite{Anderson:1991qr,Pastor-Satorras:2001fl,Bond:2012xy}. Variations of the benchmark susceptible-infected-susceptible (SIS) and susceptible-infected-recovered (SIR) models have provided valuable insights into the nature of spreading mechanisms, the dynamics of outbreaks, and the viability of containment protocols~\cite{Colizza:2007aa,Vespignani:2012,Pastor-Satorras:2015qf}. The inclusion of real-life contact and mobility patterns has yielded astonishingly accurate results, prompting the use of epidemic models as real-time predictive tools~\cite{Colizza:2006pp,Balcan22122009,Tizzoni:2012}.

The canonical modeling scheme for contact-based contagion~\cite{Anderson:1991qr,Pastor-Satorras:2015qf} assumes Markov processes and isolated transmissions. Markov processes are memoryless, which translates into exponentially distributed interevent times and enhances the mathematical tractability. The inappropriateness of this approximation, however, is widely supported by empirical evidence, such as the peaked distributions of infection periods of numerous diseases~\cite{Streftaris:2012,Chowell:2014aa} or the bursty human activity patterns in social networks, well described by heavy-tailed distributions~\cite{Oliveira:2005fk,Gonzalez:2008fk}. On the other hand, assuming isolated transmissions leads to infection channels that are not influenced by their local environment. Consequently, the infection likelihood can be written as the sum of statistically independent exposures. Nevertheless, there is evidence supporting the existence of more complex, nondyadic mechanisms, e.g., in fungal and bacterial pathogen colonization~\cite{Ben-Jacob:1994,Donabedian:2003}, and social contagion~\cite{Castellano:2008,Centola:2010,Hodas:2014}.

In recent years, an important amount of research has focused on overcoming these modeling limitations. Regarding memory, a wide array of modifications has been analyzed, such as two-step infection models~\cite{Choi:2017a,Choi:2017b}, nonexponential distributions~\cite{Van-Mieghem:2013db,Kiss:2015,Starnini:2017}, and  time-varying transmission probabilities~\cite{Dodds:2004,Lu:2011}. Conversely, a plethora of complex contagion schemes has been proposed to mediate the assumption of independent transmissions. Examples include correlated, nonlinear transmission channels~\cite{Liu:1987,Boguna:2014}, extended neighborhood effects~\cite{Perez-Reche:2011,Gomez-Gardenes:2016}, and deterministic threshold models~\cite{Watts:2002,Centola:2007}. So far, not much research has focused on tackling both assumptions simultaneously, and little is know about how these two features interact. Such a combined approach is of particular interest for contagion phenomena that include a social component, such as awareness and vaccination campaigns \cite{Granell:2013,Fu:2017}, or the spread of noncommunicable diseases (e.g., obesity, anxiety, and substance abuse)~\cite{Christakis:2007,Christakis:2008}.

In this work, we develop an infection mechanism that is equipped with memory of past exposures to multiple infectious sources. A notion of social reinforcement/inhibition arises organically, and the concepts of non-Markovian dynamics and complex contagion become intrinsically coupled. We obtain analytic results for the SIS model and perform extensive stochastic simulations in random degree-regular networks. Our analysis reveals a sophisticated interplay between two memory modes, displayed by a collective memory loss and the dislocation of the critical point into two phase transitions. An intermediate region emerges where the system is either excitable or bistable, exhibiting fundamentally distinct behaviors compared to the typical healthy and endemic phases. Additionally, the transition to the endemic phase becomes hybrid, showing both continuous and discontinuous properties.

\section{miccSIS model}
Epidemic-like models are employed for a variety of dynamics, such as opinion formation, rumor spreading, and innovation adoption. Although we use the original disease-specific terminology throughout this work, the scope of our analysis extends to all of these fields.

\subsection{General framework}
The memory-induced complex contagion susceptible-infected-susceptible (miccSIS) model describes a population of agents that can be either susceptible (healthy) or infected (infectious), and is embedded on an undirected, unweighted network. The contact network is encoded in the adjacency matrix, which is nonspatial (it carries no information about the agents' physical position) and static (it remains fixed over time).

Infected nodes have a constant infectivity rate, $\upsilon$, and continuously spread doses of contagion towards their entire neighborhood. They target all of their neighbors equally, transmitting pathogen along each edge at constant rate~$\upsilon$. Susceptible nodes collect these toxins from all their neighbors, amassing a total viral load~$\kappa$, and transition to the infected state with probability ${\psi^*_\text{inf}(\kappa)\text{d}\kappa}$, where ${\psi^*_\text{inf}(\kappa)}$ is the infection probability density. Infected nodes are unaffected by the toxins (their viral load does not increase) and recover spontaneously after a random time~$t$, with interevent time distribution~${\psi_\text{rec}(t)}$. At recovery, their viral load is completely erased and they become susceptible once again.

As in the standard SIS model, susceptible nodes whose nearest neighborhood is completely healthy cannot become infected. Since no active processes are associated to their state, they are irrelevant for the immediate evolution of the system. However, these inactive nodes play a crucial role in the long-term dynamics of the miccSIS model, therefore we assign them to an additional compartment, which we call dormant. A dormant node transitions to susceptible as soon as one of its neighbors becomes infected. Conversely, when the last infected neighbor of a susceptible node recovers, the latter transitions to the dormant state. At this point, the viral load it had previously amassed starts to deteriorate with relaxation time~$\zeta$, modeling its long-term memory. This feature mimics  the restoring of an individual's immune system, or the gradual loss of interest of an opinion, idea, or trend.

\begin{figure}[t!]
\includegraphics[width=0.8\columnwidth]{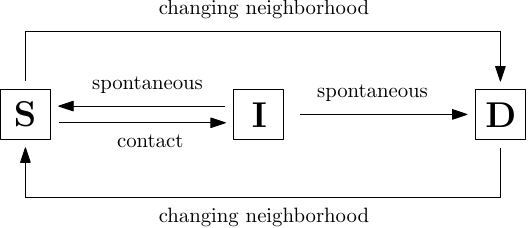}
\caption{Schematic overview of transitions in the miccSIS model.}
\label{fig:scheme}
\end{figure}

In summary, infected (I) agents spread pathogen to all their neighbors and recover spontaneously. While susceptible (S) agents have at least one infected neighbor and continuously accumulate viral load, dormant (D) agents have a fully healthy neighborhood and cannot become infected.  There are two types of active processes which entail one or possibly more transitions (see Fig.~\ref{fig:scheme}):
\begin{itemize}
\item Infection of susceptible agent~$j$.  Agent $j$ transitions from susceptible to infected. Additionally, all of $j$'s neighbors that were dormant transition to susceptible (and resume their accumulation of viral load).
\item Recovery of infected agent~$j$. If all of $j$'s neighbors are healthy, $j$ transitions from infected to dormant. If at least one of $j$'s neighbors is infected, $j$ transitions from infected to susceptible. Additionally, all of $j$'s neighbors that were susceptible and had only one infected neighbor (i.e., agent~$j$) transition to dormant (and their viral load starts to decay).
\end{itemize}
Infected agents are unaffected by the viral load, and ignore any new doses received from their infected neighbors. When an infected agent recovers it erases all the previously amassed viral load.

Overall, the system's evolution is determined by a set of discrete, stochastic processes, an infection for each susceptible node and a recovery for each infected node. All these processes are statistically independent, which enables the use of the generalized non-Markovian Gillespie algorithm~\cite{Boguna:2014}, capable of simulating memoryful dynamics in continuous time. A key ingredient of this algorithm is the instantaneous hazard rate of an active process, ${\omega(t)}$, obtained from its interevent time distribution, ${\psi(t)}$, and corresponding survival probability, ${\Psi(t)=\int_t^\infty\psi(t')\text{d}t'}$, as ${\omega(t)=\psi(t)/\Psi(t)}$. In short, $\omega(t)$ measures the probability per unit of time that the corresponding event takes place between $t$ and ${t+\text{d}t}$~\cite{cox:1970}. For a Poisson point process, the interevent time distribution is exponential and the hazard rate is, therefore, constant. In general, interevent time distributions decaying slower (respectively, faster) than exponential lead to asymptotically decreasing (increasing) hazard rates.

While recoveries are readily incorporated into this framework, $\omega_\text{rec}(t)=\psi_\text{rec}(t)/\Psi_\text{rec}(t)$, infections require some additional attention. Since the activity in a susceptible node's neighborhood varies over time, its instantaneous amassment rate, $\tilde{\upsilon}(t)$, is not constant. Then, the instantaneous hazard rate for infections is
\begin{equation}
\omega_\text{inf}(t)=\tilde{\upsilon}(t)\frac{\psi^*_\text{inf}(\kappa(t))}{\Psi^*_\text{inf}(\kappa(t))}\ ,
\end{equation}
with $\kappa(t)$ the accumulated viral load at time $t$ (see Appendix \ref{sec:appnmga} for a detailed derivation).

\subsection{Parameter selection}
In general, the infectivity rate, $\upsilon$, the relaxation time, $\zeta$, the infection probability density, $\psi^*_{\text{inf}}$, and the recovery interevent time distribution, $\psi_{\text{rec}}$, may vary from node to node. For example, one could model distinct age groups by segregating the population and assigning different values of the parameters to each subpopulation. Notwithstanding, in order to eliminate the effects of node heterogeneities, in the present work we use the same~$\upsilon$, $\zeta$, ${\psi^*_\text{inf}}$, and ${\psi_\text{rec}}$ for all nodes. For this same reason, we limit our analysis to random degree-regular networks, particularly with degree~$k=4$.

For infections we select the versatile Weibull distribution, with shape parameter~$\alpha$ and scale parameter~$\mu$,
\begin{equation}
\psi^*_\text{inf}(\kappa)=\alpha\mu^\alpha\kappa^{\alpha-1}e^{-(\mu\kappa)^\alpha}\ .
\end{equation}
When ${\alpha>1}$ it presents a peak, resembling a bell curve, $\alpha=1$ corresponds to a Poisson distribution, and for $\alpha<1$ it has power-law-like fat tails. The corresponding instantaneous hazard rate is
\begin{equation}
\omega_\text{inf}(t)=\upsilon\alpha\mu^\alpha z(t)\left[\kappa(t)\right]^{\alpha-1}\ ,\label{eq:hazardrate}
\end{equation}
with $z(t)$ the number of infected neighbors at time~$t$ (see Appendix \ref{sec:appnmga} for a detailed derivation). With ${\alpha>1}$ (respectively, ${\alpha<1}$), ${\omega_\text{inf}(t)}$ increases (decreases) monotonically with~${\kappa(t)}$.

Notice that when ${\alpha=1}$ we recover the customary expression of the standard SIS model, ${\omega_\text{inf}(t)\sim z(t)}$. For ${\alpha\neq1}$, however, Eq.~\eqref{eq:hazardrate} cannot be written as a linear superposition of independent transmission channels. In the particular case of $\alpha=2$, for example, the hazard rate includes a quadratic term, $\omega_\text{inf}(t)=az(t)+b[z(t)]^2$.
Hence, the agents' memory induces a complex contagion scheme, even though the model does not explicitly incorporate any social reinforcement/inhibition mechanisms (see Appendix \ref{sec:appcontagion} for a detailed discussion). To further isolate the effects of the modified infection mechanism, we treat recoveries as Poisson processes with rate~$\eta$, i.e., with constant hazard rate ${\omega_\text{rec}(t)=\eta}$.

We define the effective spreading ratio, $\lambda$, as the average time required to recover over the average viral load needed to become infected, nondimensionalized by the infectivity rate, ${\lambda=\upsilon\langle t_\text{rec}\rangle/\langle \kappa_\text{inf}\rangle}$. With the selected distributions we find an expression for the scale parameter, ${\mu=\eta\lambda(\alpha\upsilon)^{-1}\Gamma(\alpha^{-1})}$, with $\Gamma$ the gamma function. Notice that, once again, $\alpha=1$ recovers the customary expression of the standard SIS model, ${\lambda=\upsilon\mu/\eta}$ (with infectious rate ${\upsilon\mu}$ per infected neighbor). As for the relaxation time of dormant nodes' viral load, we consider the limit cases of instantaneous decay, ${\zeta=0}$, and perpetual accumulation, ${\zeta=\infty}$.

\begin{figure}
\includegraphics[width=\columnwidth]{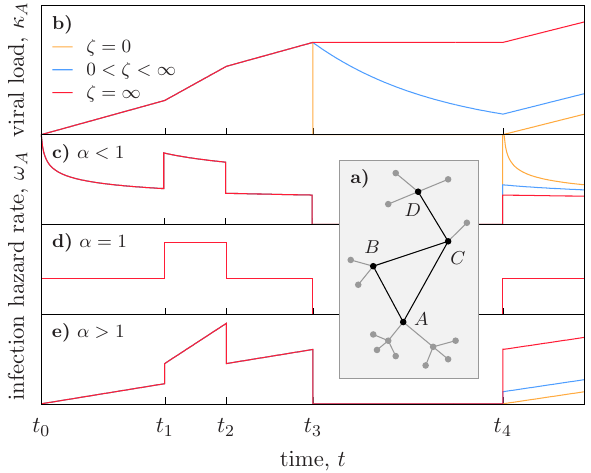}
\caption{(a) Small system considered in example. (b) Evolution of node $A$'s viral load. In the interval ${t\in[t_3,t_4]}$ node $A$ is dormant and its viral load decays instantly (orange), at a finite nonvanishing rate (blue), or accumulates perpetually (red). (c)-(e) Node $A$'s corresponding instantaneous infection rates, for (c) ${\alpha<1}$, (d) ${\alpha=1}$, and (e) ${\alpha>1}$, evaluated using Eq.~\eqref{eq:hazardrate}. In the interval ${t\in[t_3,t_4]}$ node $A$'s neighborhood is fully healthy and it cannot become infected (${\omega_A=0}$).}
\label{fig:loadrates}
\end{figure}

For illustrative purposes, consider the system depicted in Fig.~\ref{fig:loadrates}(a), where all nodes are initially healthy except for~$D$. Node $C$ becomes infected at time $t_0$ and subsequently infects $B$ at~$t_1$. During the interval ${t\in[t_0,t_1]}$, node $A$'s viral load, $\kappa_A$, grows with rate~$\upsilon$, but from $t_1$ onwards it will increase with rate~$2\upsilon$. At~$t_2$, node $C$ recovers and $\kappa_A$ reduces its accumulation rate back to~$\upsilon$, and when $B$ recovers at~$t_3$, $\kappa_A$ starts to decay with relaxation time~$\zeta$. Finally, $C$ becomes infected once again at $t_4$ and $\kappa_A$ resumes its growth at rate~$\upsilon$. Figs.~\ref{fig:loadrates}(b-e) show the evolution of $\kappa_A$ and $\omega_A$ for various $\zeta$ and~$\alpha$. Hereafter we use temporal units such that $\eta=1$ and, without loss of generality, set~${\upsilon=1}$.

\section{Short-term memory}
\subsection{Analytics}
We begin our analysis for ${\zeta=0}$, with dormant nodes instantly erasing their viral load. Thus, when the outbreak reenters their neighborhood, they become susceptible starting afresh (with ${\kappa=0}$). Hence, the only memory effect present is during the infection period, which we interpret as a short-term memory mode.

Each node is uniquely defined by its state, infected or healthy, and its accumulated viral load. The state of a node only changes with the transitions i) infected to healthy, and ii) susceptible to infected. When a node transitions between susceptible and dormant, its state remains unaltered. On the other hand, the viral load i) is erased instantly when an infected node recovers, ii) increases proportionally to the number of infected neighbors while a node is susceptible, and iii) is erased instantly when a susceptible node becomes dormant.

In the Supplemental Material (SM), we derive exact stochastic evolution equations for both the state and accumulated viral load of each node. Applying a mean-field approximation, these equations can be reduced to a single expression for the late-time prevalence, $\rho$, the average fraction of nodes that are infected in the steady-state (see SM for a detailed derivation). If all nodes have the same degree~$k$, this equation reads
\begin{equation}
-\rho+ak\rho(1-\rho)-bk\rho(1-\rho)^k=0\ ,\label{eq:analitic}
\end{equation}
with ${a>0}$. Roughly speaking, the first term corresponds to the recovery of infected nodes, the second term to the infection of susceptible nodes, and the third is related to susceptible nodes becoming dormant. The shape factor of the infection probability, $\alpha$, and the effective spreading ratio, $\lambda$, are encapsulated in~$a$ and~$b$, which become constant parameters when $\rho \approx 0$.

Equation~\eqref{eq:analitic} uncovers the existence of an epidemic threshold, where the population transitions from a healthy, pathogen-free state to an endemic, disease burdened one. Linear stability analysis reveals that the healthy phase, $\rho=0$, becomes unstable at ${a-b-1/k=0}$, yielding a phase transition at the critical point ${a_\text{c}=b+1/k}$. Moreover, the nature of the transition changes at ${a-bk=0}$, yielding a tricritical point at ${a_\text{tc}=1/(k-1)}$, ${b_\text{tc}=1/k(k-1)}$. Then the steady state is endemic for ${a>a_\text{c}}$, and the phase transition is continuous for ${b<b_\text{tc}}$ and discontinuous for ${b>b_\text{tc}}$. Finally, the prevalence of the endemic phase scales as ${\rho\propto(a-a_\text{c})^\beta}$, with ${\beta_\text{c}=1}$ when the transition is continuous, and ${\beta_\text{tc}=1/2}$ at the tricritical point.

\begin{figure}
\includegraphics[width=\columnwidth]{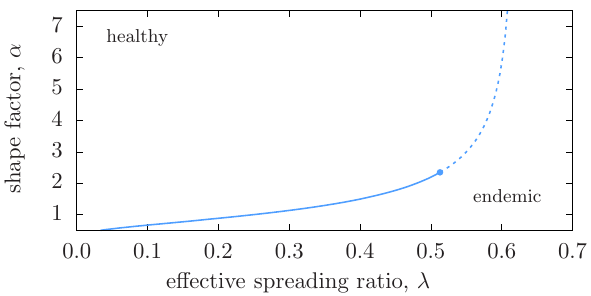}
\caption{Phase diagram for ${\zeta=0}$. The solid lines indicates a continuous phase transition, the dashed line indicates a discontinuous phase transition, and the dot marks the tricritical point, ${\alpha_\text{tc}\approx2.348}$, ${\lambda_\text{tc}\approx0.513}$.}
\label{fig:phasediag}
\end{figure}

Figure~\ref{fig:phasediag} shows the phase diagram in terms of the original parameters $\alpha$ and~$\lambda$. We observe that the critical point initially increases monotonically with~$\alpha$, but afterwards saturates for ${\alpha\rightarrow\infty}$. This result is consistent with the possibly largest epidemic threshold reported in~\cite{Liu:2018}. On the other hand, the transition to endemicity is discontinuous for ${\alpha>\alpha_\text{tc}\approx2.348}$. Recall that the infection probability density presents a peak for ${\alpha>1}$, and, in fact, tends towards a delta function in the limit ${\alpha\rightarrow\infty}$. Thus, a node requires a quasi deterministic amount of viral load to become infected, mimicking a threshold model, which commonly exhibits a discontinuous phase transition~\cite{Min:2018}.

\subsection{Simulations}
In order to verify these analytic findings, we perform extensive stochastic simulations. We begin by exploring the position of the critical point, ${\lambda_\text{c}}$, that separates an absorbing (healthy) phase (${\lambda<\lambda_\text{c}}$) from an active (endemic) one (${\lambda>\lambda_\text{c}}$). The simulations start well into the active phase with a fully infected population, and quasistatically decrease the control parameter, $\lambda$, until finite-size fluctuations trap the system in the absorbing state. We sample the late-time prevalence, ${\rho_\text{st}=\lim_{t\rightarrow\infty}N_\text{I}(t)/N}$, of $10^4$~states, time-averaged over various trajectories (see SM for simulation details).

\begin{figure}
\includegraphics[width=\columnwidth]{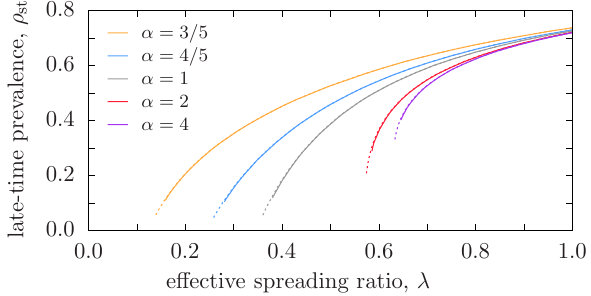}
\caption{Late-time prevalence of the active steady state for networks of size ${N=10^3}$ (solid) and ${N=10^4}$ (dashed), with ${\zeta=0}$. Infection probability distributions with ${\alpha=3/5}$ (orange), ${\alpha=4/5}$ (blue), ${\alpha=1}$ (grey), ${\alpha=2}$ (red), and ${\alpha=4}$ (purple). Uncertainty bars not appreciable at this scale.}
\label{fig:prevalence0}
\end{figure}

As shown in Fig.~\ref{fig:prevalence0}, ${\lambda_\text{c}}$ indeed increases with~$\alpha$, and saturates for large values of~$\alpha$. Moreover, for ${\alpha<1}$ the approach to the critical point is very similar to the standard SIS (${\alpha=1}$), consistent with a continuous phase transition with exponent ${\beta=1}$. On the other hand, for ${\alpha=4}$ the curves terminate at a remarkably high prevalence, consistent with a discontinuous phase transition. Finally, the curves for ${\alpha=2}$ also terminate at a rather high prevalence, which deviates from the analytic prediction. Nonetheless, since the apparent discontinuity decreases with the system size, this observation is most likely related to finite-size effects.

We complement these results with the analysis of patient zero scenarios, the arrival of an infected agent in a previously unaffected population. We employ the lifespan method~\cite{Mata:2015}, which simulates outbreaks starting from a single infected node. Each single-seed realization is characterized by its coverage, $K$, defined as the number of distinct nodes that have become infected at least once, and can be either finite or endemic. While the former return to the absorbing state, the latter evolve towards an active steady state. In finite systems, we introduce a coverage threshold, ${K_\text{th}=c_\text{th}N}$, with ${0<c_\text{th}<1}$. A realization is declared endemic whenever its coverage reaches the threshold, and those that terminate without surpassing it are considered finite. As reported in~\cite{Mata:2015}, the value of $c_{\text{th}}$ does not modify the qualitative results. Hereafter, we use ${c_\text{th}=0.75}$.

\begin{figure}
\includegraphics[width=\columnwidth]{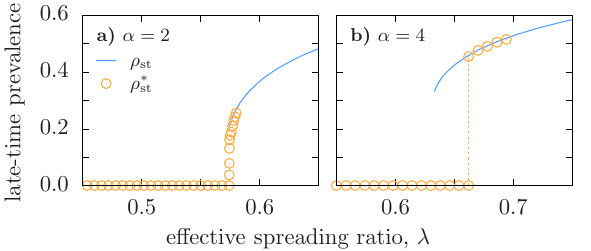}
\caption{Late-time prevalence for quasistatic (blue curve) and single-seed (orange circles) simulations with ${\zeta=0}$ in a network of ${N=10^4}$ nodes, for (a) ${\alpha=2}$ and (b) ${\alpha=4}$. Uncertainty bars comparable to symbol size.}
\label{fig:ls-0-greater}
\end{figure}

For a fixed value of $\lambda$ we run $10^4$~realizations, each starting with a single, randomly chosen infected node, and a system cleared of all viral load. If an outbreak becomes endemic, we extend the simulation until it reaches the steady-state, and then measure the late-time prevalence,~$\rho_\text{st}^*$ (see SM for simulation details). The results are shown in Figs.~\ref{fig:ls-0-greater}(a,b), which include the previously computed ${\rho_\text{st}}$. While $\rho_\text{st}^*$ shows a discontinuous jump for $\alpha=4$, with $\alpha=2$ it grows continuously and coincides with $\rho_\text{st}$, supporting our analytical findings. Moreover, since ${\alpha=2}$ is close to tricritical point, a cross-over effect towards the exponent ${\beta_\text{tc}}$ is expected, explaining the rather steep approach towards the critical point.

 Overall, these analytic and simulated results indicate that the system's macroscopic properties are drastically affected by the microscopic details of the infection mechanism. In particular, the critical point that separates the healthy from the endemic phase grows with $\alpha$, and the nature of the phase transition changes from continuous to discontinuous at the tricrital point $\alpha_\text{tc}$.

\section{Long-term memory}
Next, we consider the case ${\zeta=\infty}$, where a dormant node's viral load remains frozen until the outbreak revisits its neighborhood.  Besides the short-term memory present during the infection period, nodes now possess an additional long-term memory mode that is capable of connecting very distant temporal points, causing the system to evolve in a highly nonlinear manner.

As before, the state of a node changes with the transitions i) infected to healthy, and ii) susceptible to infected. On the other hand, the viral load i) is instantly erased when an infected node recovers, ii) increases proportionally to the number of infected neighbors while a node is susceptible, and iii) remains constant while the node is dormant. We can write an equation for the state of node~$i$ (${n_i=1}$ if its infected, ${n_i=0}$ if its healthy), and obtain the expected value in the steady-state (see SM for a detailed derivation). The resulting equation reads
\begin{equation}
\langle n_i\rangle=\lambda\langle(1-n_i)z_i\rangle\ ,\label{eq:rho8}
\end{equation}
with ${z_i=\sum_ja_{ij}n_j}$ the number of infected neighbors, and $a_{ij}$ the elements of the adjacency matrix. Note that this equation is derived without implementing a mean-field approximation.

\begin{figure}
\includegraphics[width=\columnwidth]{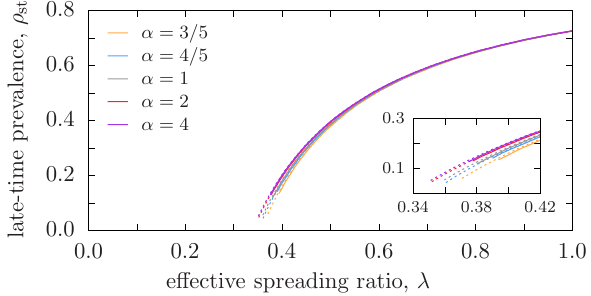}
\caption{Late-time prevalence of the active steady state for networks of size ${N=10^3}$ (solid) and ${N=10^4}$ (dashed), with ${\zeta=\infty}$. Infection probability distributions with ${\alpha=3/5}$ (orange), ${\alpha=4/5}$ (blue), ${\alpha=1}$ (grey), ${\alpha=2}$ (red), and ${\alpha=4}$ (purple). Uncertainty bars not appreciable at this scale.}
\label{fig:prevalenceA}
\end{figure}

Surprisingly, Eq.~\eqref{eq:rho8} is identical to the first-order equation of the standard SIS model~\cite{Pastor-Satorras:2015qf} and, therefore, both the miccSIS and SIS models have the same mean field description. Thus, whenever the standard SIS dynamics is well described by its mean field approximation, as is the case for random degree-regular networks, the miccSIS dynamics is also aptly described by the same mean field equation. This equivalence demonstrates that the late-time prevalence is independent of~$\alpha$, and identical to the Markovian model. This result is verified by stochastic simulations, shown in Fig.~\ref{fig:prevalenceA}. Indeed we observe that the late-time prevalence curves coincide for all values of~$\alpha$. A small deviation occurs in the critical region (see inset of Fig.~\ref{fig:prevalenceA}), which is caused by finite size fluctuations.

Compared to the short-term memory mode, for ${\alpha>1}$ (respectively, ${\alpha<1}$) the endemic phase is enlarged (shrunken) by the long-term mode. This phenomenon is explained by the monotonically increasing (decreasing) infection rate. When the outbreak revisits a dormant node's neighborhood, its previously accumulated viral load facilitates (hinders) reinfection, enabling (preventing) the outbreak to remain active in a wider range of~$\lambda$. These results reveal that the additional long-term memory completely suppresses the effects of the short-term mode. Specifically, it causes individuals with virtually infinite memory to behave, on the aggregate, as if they had no memory at all. This collective memory loss consequently renders the system's macroscopic state unable to distinguish between agents' microscopic properties.

In order to elucidate these findings, we proceed with the analysis of patient zero scenarios, where an infected agent arrives in a previously unaffected population. For a fixed value of $\lambda$ we run $10^4$~realizations, each starting with a single, randomly chosen infected node, and a system cleared of all viral load. We measure the average coverage fraction, ${\bar{c}=\langle K\rangle/N}$, and the probability that a realization surpasses the coverage threshold, $P_1$, which serves as a proxy for the true endemic probability, $P_{\infty}$, the probability, in the thermodynamic limit, that an outbreak becomes endemic (see SM for simulation details).

\begin{figure}
\includegraphics[width=\columnwidth]{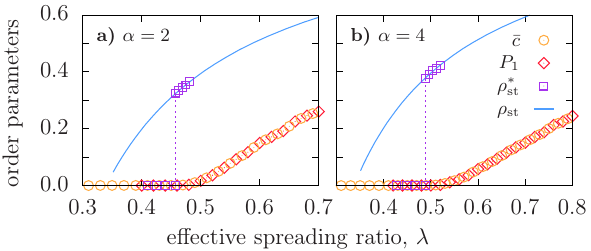}
\caption{Average coverage (orange circles), endemic probability (red diamonds), and late-time prevalences (purple squares and blue curve) with ${\zeta=\infty}$ in a network of ${N=10^4}$ nodes, for (a) ${\alpha=2}$ and (b) ${\alpha=4}$. Uncertainty bars not appreciable at this scale.}
\label{fig:ls-A-greater}
\end{figure}

\subsection{Bistability}
We first analyze $\alpha>1$, for which the infection probability presents a peak and the instantaneous infection rate increases monotonically with the accumulated viral load. The patient zero results are shown in Figs.~\ref{fig:ls-A-greater}(a,b), which include the previously computed ${\rho_\text{st}}$, and the late-time prevalence of single-seed outbreaks that are able to become endemic, ${\rho_\text{st}^*}$ (see SM for simulation details). We find that the average coverage, $\bar{c}$, and the endemic probability, $P_1$, coincide for all values of $\lambda$ and present a continuous phase transition at ${\lambda_\text{c}(\bar{c})}$. However, this point is notably larger than the critical point of the late-time prevalence, ${\lambda_\text{c}(\rho_\text{st})}$. While $\rho_\text{st}$ presents a continuous phase transition, $\rho_\text{st}^*$ exhibits a discontinuous phase transition at ${\lambda_\text{c}(\rho_\text{st}^*)=\lambda_\text{c}(\bar{c})}$. As expected, the two prevalence curves overlap after the abrupt jump.

This evidences the existence of an intermediate region ${\lambda\in[\lambda_\text{c}(\rho_\text{st}), \lambda_\text{c}(\bar{c})]}$ where all single-seed outbreaks return to the absorbing state, whereas fully infected populations evolve towards an active steady state. The key ingredient to understand this phenomenon is the environment of frozen viral load. During the simulations that measure the late-time prevalence, the viral loads are well thermalized, enabling the outbreak to remain in an active state. Conversely, this environment is deficient in single-seed outbreaks, as the system has not yet reached its steady state. Hence, outbreaks are unable to produce sufficient new infections and rapidly become trapped in the absorbing state. 

These results indicate that the system displays two attractors in this intermediate region. Then, for ${\zeta=\infty}$ and ${\alpha>1}$ the system's phase diagram exhibits an additional bistable phase, that separates the usual healthy and endemic phases. The associated hysteresis loop, however, has a rather exotic nature: although its lower branch presents the expected discontinuity, the upper branch connects the two attractors in a continuous manner. This contrasts with previous findings of bistability, were the hysteresis loop is bounded by two discontinuities~\cite{Dodds:2004,Gomez-Gardenes:2016,Chen:2018}. Moreover, the transition to full endemicity is hybrid~\cite{Cai:2015}: the endemic probability grows continuously at $\lambda_\text{c}(\rho_\text{st}^*)$, but the late-time prevalence jumps discontinuously.

\begin{figure}
\includegraphics[width=\columnwidth]{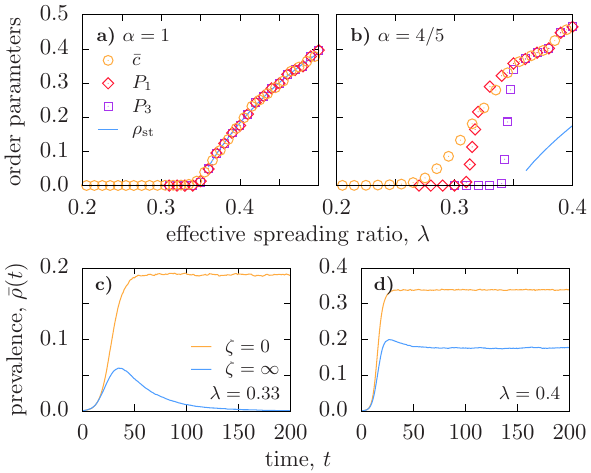}
\caption{(Top) Average coverage (orange circles), endemic probabilities (red diamonds and purple squares), and late-time prevalence (blue curve) with ${\zeta=\infty}$ in a network of ${N=10^4}$ nodes, for (a) ${\alpha=1}$ and (b) ${\alpha=4/5}$. Uncertainty bars not appreciable at this scale. (Bottom) Evolution of single-seed outbreaks that reach the coverage threshold once for ${\alpha=4/5}$, with $\zeta=0$ (orange) and ${\zeta=\infty}$ (blue) in a network of ${N=10^4}$ nodes, for (c) ${\lambda=0.33}$ and (d) ${\lambda=0.4}$, averaged over $10^2$~trajectories. Uncertainty bars not appreciable at this scale.}
\label{fig:ls-A-smaller}
\end{figure}

\subsection{Excitability}
Finally, we study $\alpha<1$, for which the infection probability presents power-law-like fat tails an the instantaneous infection rate decreases monotonically with the accumulated viral load. In Figs.~{\ref{fig:ls-A-smaller}(a,b)} we show the patient zero analysis for the standard SIS (${\alpha=1}$) and broad-tailed infection distributions (${\alpha<1}$). Here, we additionally compute~$P_3$, the probability that a single-seed outbreak reaches the coverage threshold three times (see SM for simulation details). With $\alpha=1$, $P_1$ and $P_3$ are practically identical, indicating that an outbreak that surpasses the coverage threshold once remains active long enough to surpass the threshold two more times. Thus $P_1$ is an adequate proxy for the true endemic probability, $P_{\infty}$, which additionally coincides with $\bar{c}$ and $\rho_\text{st}$.

For ${\alpha<1}$ the situation is quite different. Firstly, the average coverage starts growing continuously at ${\lambda_\text{c}(\bar{c})}$, when all other order parameters are still identically zero. Additionally, the transition point of $P_1$ is significantly lower than that of~$P_3$. Thus, there is a wide interval where all outbreaks that surpass the threshold once eventually terminate in the absorbing state, evidencing the inadequateness of $P_1$ as a measure of the true endemic probability. The inflection point of $P_3$ is much closer to the transition point of ${\rho_\text{st}}$, which suggests that the critical point of the endemic probability ($P_\infty$, the probability to surpass the threshold an infinite amount of times) coincides with ${\lambda_\text{c}(\rho_\text{st})}$. Beyond this point, $P_{\infty}$ is expected to coincide with $\bar{c}$, indicating that the endemic probability presents a discontinuous phase transition. Then, the transition to full endemicity is again hybrid. Nonetheless, here the late-time prevalence grows continuously, while the endemic probability jumps discontinuously.

In this case, we find an intermediate region ${\lambda\in[\lambda_\text{c}(\bar{c}),\lambda_\text{c}(\rho_\text{st})]}$ where outbreaks are unable to become endemic (${P_{\infty}=0}$) but affect a macroscopic fraction of the population (${c>0}$). In Figs.~\ref{fig:ls-A-smaller}(c,d) we show the prevalence, ${\bar{\rho}(t)=\langle N_\text{I}(t)\rangle/N}$, of realizations that surpass the coverage threshold once, for ${\lambda=0.33}$ and ${\lambda=0.4}$. We include the results for ${\zeta=0}$ for comparison (see SM for simulation details). Both values of $\lambda$ are located in the active phase of the short-term memory mode (${\zeta=0}$), and so the endemic realizations converge monotonically towards their active steady states. For the long-term memory mode (${\zeta=\infty}$), ${\lambda=0.33}$ is located in the intermediate region. In this case, we observe that outbreaks grow up to a maximum, after which their prevalence gradually diminishes until they reach the absorbing state.  This behavior is typically observed in SIR-like dynamics and is reminiscent of excitable media. In the endemic phase (${\lambda=0.4}$) the outbreaks continue presenting a peak, but afterwards relax towards an active steady state.

In conclusion, for ${\zeta=\infty}$ and ${\alpha<1}$ the usual healthy and endemic phases are separated by an additional excitable phase. This excitable behavior is again a consequence of the environment of frozen viral load. Independently of~$\zeta$, an outbreak starts from a single infected node in a population cleared of viral load. Then it initially evolves as if the agents only had the short-term memory mode (clearly appreciable for $t\in[0,25]$ in Fig.~{\ref{fig:ls-A-smaller}(c)}), rapidly achieving a large coverage. When the outbreak revisits a previously affected area, the long-term memory mode is activated and the frozen viral load impedes new infections. Thus, dormant nodes are effectively removed from the dynamic, impede the outbreak to grow, and eventually cause its extinction. To the extend of our knowledge, this excitable behavior has not been previously reported in comparable SIS-like models.

\section{Conclusions}
All these results show a crucial feature of agents that possess the long-term memory mode. Focusing only on the late-time prevalence of fully infected populations provides little insight about the system's constituents. Nevertheless, widely distinct and clearly distinctive behaviors appear with the analysis of patient zero scenarios. Moreover, a common effect of agents' memory is the breaking of the symmetry between the order parameters $\bar{c}$, $P$, and~${\rho_\text{st}}$. If agents are memoryless, all three order parameters are completely identical. This symmetry is broken when agents possess a long-term memory mode, and the critical points become dissociated. The system first transitions from the healthy phase to an either bistable or excitable intermediate regime, followed by a hybrid transition to the endemic phase. This differs from a double phase transition, where the same order parameter undergoes two consecutive phase transitions, a phenomenon usually associated to node and/or topological heterogeneities~\cite{Dodds:2004,Chen:2018,Colomer:2014,Allard:2017}.

The analysis of our stylized, yet feature-rich model evidences a crucial role of non-Markovianity in the spread of epidemic outbreaks. In particular, the agents' memory range dramatically impacts the outbreak of newly introduced pathogens. This topic is currently a very active field of epidemic modeling, with applications ranging from the appearance of exotic diseases, to the dissemination of fake news on social media.

Nonetheless, further research is necessary. A thorough study of finite, non-vanishing relaxation times of the viral load of dormant nodes can aid in further elucidating the interplay between memory modes, and its effects on the system's properties. Furthermore, the inclusion of nontrivial contact networks will supply renewed insight on the relevance of microscopic mechanisms and topological properties in dynamical processes on networks.

\begin{acknowledgments}
We acknowledge support from a James S. McDonnell Foundation Scholar Award in Complex Systems; the ICREA Academia prize, funded by the Generalitat de Catalunya; and Ministerio de Econom\'{\i}a y Competitividad of Spain, project no. FIS2016-76830-C2-2-P (AEI/FEDER, UE); X.~R.~H. acknowledges support from the Ministerio de Educaci\'{o}n, Cultura y Deporte of Spain, scholarship no. FPU16/05751.
\end{acknowledgments}

\appendix

\section{Non-Markovian Gillespie algorithm}\label{sec:appnmga}
\subsection{General framework}
Here, we summarize the derivation reported in~\cite{Boguna:2014}. Consider a set of $M$ statistically independent, discrete, stochastic processes, each with an interevent time distribution ${\psi_j(t)}$ and corresponding survival probability ${\Psi_j(t)=\int_t^\infty\psi_j(t')\text{d}t'}$. At a certain moment in time~$t_0$, process $j$ has been active for $t_j$ units of time. Let ${\phi(\tau,i|\{t_k\})\text{d}\tau}$ denote the joint probability that the next-occurring event takes place in the interval ${t\in[t_0+\tau, t_0+\tau+\text{d}\tau]}$ and corresponds to process~$i$, conditioned by the set of elapsed times~$\{t_k\}$. This probability density can be expressed as
\begin{equation}
\phi(\tau,i|\{t_k\})=\frac{\psi_i(t_i+\tau)}{\Psi_i(t_i+\tau)}\Phi(\tau|\{t_k\})\ ,
\end{equation}
where
\begin{equation}
\Phi(\tau|\{t_k\})=\prod_{j=1}^M\frac{\Psi_j(t_j+\tau)}{\Psi_j(t_j)}
\end{equation}
is the survival probability of~$\tau$, i.e., the conditional probability that no event takes place before ${t_0+\tau}$. Then the probability that the next event takes place in the interval ${t\in[t_0,t_0+\tau]}$ is
\begin{equation}
\Xi(\tau|\{t_k\})=1-\Phi(\tau|\{t_k\})\ .\label{eq:sampletau}
\end{equation}
Once the interval $\tau$ is known, the probability that the next-occurring event corresponds to process $i$ is given by
\begin{equation}
\Pi(i|\tau,\{t_k\})=\frac{\omega_i(t_i+\tau)}{\sum_{j=1}^M\omega_j(t_j+\tau)}\ ,\label{eq:sampleprocess}
\end{equation}
with ${\omega_j(t)=\psi_j(t)/\Psi_j(t)}$ the instantaneous hazard rate of process~$j$.

Eqs.~\eqref{eq:sampletau} and~\eqref{eq:sampleprocess} provide an algorithm that generates statistically correct sequences of events: i)~draw the interval by solving ${\Xi(\tau|\{t_k\})=u}$, with ${u\in U(0,1)}$, ii)~increase the system time as ${t\leftarrow t+\tau}$, iii)~draw the process from the discrete distribution ${\Pi(i|\tau,\{t_k\})}$, iv)~revise the list of active processes, and v)~update the set of elapsed times as ${t_j\leftarrow t_j+\tau}$ (setting ${t_j=0}$ for newly activated processes).
\subsection{Incorporating viral loads}
Recoveries are straightforwardly incorporated into this framework, with the elapsed time~$t_j$ measuring the period since agent $j$ became infected (i.e., this occurred at ${t=t_0-t_j}$). On the other hand, infection processes require the translation of infection probability densities into interevent time distributions. Consider susceptible agent~$j$, characterized by its infection probability density ${\psi^*_j(\kappa)}$ and corresponding survival probability ${\Psi_j^*(\kappa)=\int_{\kappa}^\infty\psi^*_j(\kappa')\text{d}\kappa'}$. Its interevent time distribution, ${\psi_j(t)}$, is given by the normalization condition
\begin{equation}
\psi_j(t)\text{d}t=\psi_j^*(\kappa)\text{d}\kappa\ .
\label{eq:norm}
\end{equation}
Since the activity in $j$'s neighborhood may vary over time, the rate at which it amasses viral load is generally nonconstant. At time~$t$ it has amassed ${\kappa_j(t)}$ units of viral load and has ${z_j(t)}$ infected neighbors. If the system remains unaltered in an interval ${\text{d}t}$, node $j$ will amass an additional ${\text{d}\kappa=\tilde{\upsilon}_j\text{d}t}$, with ${\tilde{\upsilon}_j(t)=\sum_{i=1}^{z_j(t)}\upsilon_i}$ its instantaneous amassment rate. Substituting in Eq.~\eqref{eq:norm} we find 
\begin{equation}
\psi_j(t)=\tilde{\upsilon}_j(t)\psi_j^*(\kappa_j(t))\ .
\end{equation}
For the survival probability we have
\begin{equation}
\Psi_j(t)=\int_{t}^\infty\psi_j(t')\text{d}t'=\int_{\kappa_j(t)}^\infty\psi^*_j(\kappa')\text{d}\kappa'=\Psi^*_j(\kappa_j(t))\ ,
\end{equation}
which yields its instantaneous hazard rate
\begin{equation}
\omega_j(t)=\tilde{\upsilon}_j(t)\frac{\psi_j^*(\kappa_j(t))}{\Psi_j^*(\kappa_j(t))}\ .
\end{equation}
Note that we can always write ${t=t_0+\tau}$, with $t_0$ the time at which the system was last updated, and ${\tau\geq0}$. Then, the instantaneous amassment rate remains constant in the interval ${[t_0,t]}$, ${\tilde{\upsilon}_j(t)=\tilde{\upsilon}_j(t_0)}$, and ${\kappa_j(t)=\kappa_j(t_0)+\tau\tilde{\upsilon}_j(t_0)}$.

In our work, infections are governed by a Weibull distribution with shape parameter~$\alpha$ and scale parameter~$\mu$,
\begin{equation}
\psi_\text{inf}^*(\kappa)=\alpha\mu^\alpha\kappa^{\alpha-1}e^{-(\mu\kappa)^\alpha}\hspace{24pt}\Psi_\text{inf}^*(\kappa)=e^{-(\mu\kappa)^\alpha}\ ,
\end{equation}
 and recoveries by a Poisson process with rate $\eta$,
\begin{equation}
\psi_\text{rec}(t)=\eta e^{-\eta t}\hspace{24pt}\Psi_\text{rec}(t)=e^{-\eta t}\ .
\end{equation}
Since all infectors have the same infectivity rate $\upsilon$, the instantaneous amassment rate of a susceptible node $j$ becomes ${\tilde{\upsilon}_j(t)=\upsilon z_j(t)}$, with $z_j(t)$ the number of its neighbors that are infected at time $t$. So the instantaneous hazard rates for infections and recoveries are, respectively, ${\omega_\text{inf}(t)=\upsilon\alpha\mu^{\alpha}z(t)[\kappa(t)]^{\alpha-1}}$ and ${\omega_\text{rec}(t)=\eta}$.

\section{Simple and complex contagion}\label{sec:appcontagion}
\subsection{Simple contagion}
Simple contagion describes purely dyadic interactions, thus we can identify each edge that connects a healthy node with an infected one as an isolated transmission channel. Consider at time $t$ a susceptible node~$j$ and its infected neighbor~$i$, which became infected at ${t_i<t}$. The probability that node~$i$ infects node~$j$ within the interval $(t,t+\text{d}t)$ is ${\omega_{i\rightarrow j}(t|t_i)\text{d}t}$, regardless of the rest of the system. If node~$j$ has ${z_j(t)}$ infectors at time~$t$, the previous statement holds for each of them.

The total probability that node~$j$ becomes infected at time~$t$ depends on all of its incoming transmission channels. Since these are statistically independent, we can write
\begin{equation}
\omega_j(t)=\sum_{i=1}^{z_j(t)}\omega_{i\rightarrow j}(t|t_i)\ ,\label{eq:simcon}
\end{equation}
where $\omega_j(t)$ is the instantaneous hazard rate of node~$j$'s infection process (i.e., the probability per unit of time that node~$j$ becomes infected at time~$t$). Using
\begin{equation}
\Omega_j(t)=\frac{1}{z_j(t)}\sum_{i=1}^{z_j(t)}\omega_{i\rightarrow j}(t|t_i)\label{eq:ave}
\end{equation}
we can write Eq.~\eqref{eq:simcon} as
\begin{equation}
{\omega_j(t)=\Omega_j(t)z_j(t)}\ ,\label{eq:ratesis}
\end{equation}
thus the total hazard rate is proportional to the number of current infectors. If ${\omega_{i\rightarrow j}(t|t_i)=\beta}$ are constants (and homogenous for all pairs of nodes), we recover the standard SIS model with the familiar expression ${\omega_j(t)=\beta z_j(t)}$. When the transmission rates ${\omega_{i\rightarrow j}}$ are time-dependent, the dynamics has memory effects; thus, simple contagion can be non-Markovian (as in~\cite{Starnini:2017}, for example).

\subsection{Complex contagion}
When the dynamics are described by interactions that are not strictly dyadic, the contagion becomes complex. These processes usually incorporate an explicit social reinforcement or inhibition mechanism. Although the classification of complex contagion processes is yet to be formalized, they can be broadly categorized into two groups:
\begin{itemize}
\item In edge-centric approaches, one still considers the transmission channel from infected node~$i$ to susceptible node~$j$. Now, however, the transmission rate ${\omega_{i\rightarrow j}}$ is affected by the neighborhoods of~$i$ and/or~$j$ (for specific examples, see~\cite{Perez-Reche:2011,Gomez-Gardenes:2016,Liu:2017}). Considering only nearest-neighbors, the transmission rate from node~$i$ to node~$j$ at time~$t$, ${\omega_{i\rightarrow j}(t|z_i(t),z_j(t))}$, is a function of their current infected neighbors, $z_i(t)$ and $z_j(t)$. Although we can still write the total hazard rate ${\omega_j(t)}$ as in Eq.~\eqref{eq:simcon}, the instantaneous average defined in Eq.~\eqref{eq:ave} has an explicit dependence on $z_i(t)$ and $z_j(t)$. Consequently, ${\omega_j(t)}$ can be superlinear (reinforcement) or sublinear (inhibition) with the number of current infectors, ${z_j(t)}$.

\item On the other hand, node-centric approaches forgo the notion of transmission channels and directly prescribe the instantaneous hazard rate, ${\omega_j(t)}$. These usually incorporate thresholds, such as ${\omega_j(t)=\delta(T_j-z_j(t))}$~\cite{Watts:2002} or ${\omega_j(t)=z_j(t)\Theta(T_j-z_j(t))+\beta\Theta(z_j(t)-T_j)}$~\cite{OSullivan:2015}, which explicitly evidence the nonlinearity of ${\omega_j(t)}$ with ${z_j(t)}$.
\end{itemize}

\subsection{Memory-induced complex contagion}
Consider an isolated pair of nodes $i$ and $j$ in the miccSIS model. Both are healthy when node $i$ becomes infected at time $t_i$. The total amount of viral load that $j$ has amassed at time $t>t_i$ is $\kappa_j(t)=\upsilon(t-t_i)$. The instantaneous hazard rate of node $j$'s infection is
\begin{equation}
\omega_j(t)=\upsilon\alpha\mu^{\alpha}z_j(t)[\kappa_j(t)]^{\alpha-1}\ ,\label{eq:omegaj}
\end{equation}
but since $j$ has only one infected neighbor, it is given solely by the exposure to node $i$: $\omega_j(t)=\omega_{i\rightarrow j}(t|t_i)$, with
\begin{equation}
\omega_{i\rightarrow j}(t|t_i)=\upsilon\alpha\mu^\alpha[\upsilon(t-t_i)]^{\alpha-1}\ .\label{eq:omegaij}
\end{equation}

Now consider an infected node $j$ that at time $t_0$ recovers and becomes dormant (all its neighbors are healthy). At time $t$ it has a set of $\{i\}$ infected neighbors that became infected at times $\{t_i\}$, with ${t_0<t_i<t}$, ${\forall i=1,2,...,z_j(t)}$. The total amount of viral load that $j$ has amassed at time $t$ is
\begin{equation}
\kappa_j(t)=\chi(t)+\sum_{i=1}^{z_j(t)}\upsilon(t-t_i)\ ,\label{eq:kappaj}
\end{equation}
where $\chi(t)$ stores the viral load accumulated from neighbors that became infected in the interval ${(t_0, t)}$ but are currently healthy. Substituing Eq.~\eqref{eq:kappaj} in Eq.~\eqref{eq:omegaj}, we find
\begin{equation}
\omega_j(t)=\upsilon\alpha\mu^\alpha z_j(t)\left[\chi(t)+\sum_{i=1}^{z_j(t)}\upsilon(t-t_i)\right]^{\alpha-1}\ ,
\end{equation}
which, using Eq.~\eqref{eq:omegaij}, can be written for $\alpha\neq1$ as
\begin{equation}
\omega_j(t)=z_j(t)\left[(\upsilon\alpha\mu^\alpha)^{\frac{1}{\alpha-1}}\chi(t)+\sum_{i=1}^{z_j(t)}\left[\omega_{i\rightarrow j}(t|t_i)\right]^{\frac{1}{\alpha-1}}\right]^{\alpha-1}\ .\label{eq:nmomegaj}
\end{equation}
Notice that Eq.~\eqref{eq:nmomegaj} cannot be written as Eq.~\eqref{eq:simcon}. Therefore, while the exposures to infectious sources are initially described as isolated events, the agents' memory causes them to become entangled. For instance, for $\alpha=2$, Eq.~\eqref{eq:nmomegaj} can be written as
\begin{equation}
\omega_j(t)=2 \upsilon \mu \chi(t) z_j(t)+ \Omega(t)[z_j(t)]^2
\end{equation}
which has an explicit quadratic dependence on $z_j(t)$~\footnote{Notice, however, that the terms $\chi(t)$ and $\Omega(t)$ are stochastic processes that play an important role in the global dynamics.}.
The simple contagion of the standard SIS model can be recovered only with $\alpha=1$, for which Eq~\eqref{eq:omegaj} equates with Eq~\eqref{eq:ratesis}. In conclusion, the non-Markovianity of the miccSIS model induces an effective social reinforcement/inhibition even though it was not incorporated in the initial description of the model.


\begin{thebibliography}{46}%
\makeatletter
\providecommand \@ifxundefined [1]{%
 \@ifx{#1\undefined}
}%
\providecommand \@ifnum [1]{%
 \ifnum #1\expandafter \@firstoftwo
 \else \expandafter \@secondoftwo
 \fi
}%
\providecommand \@ifx [1]{%
 \ifx #1\expandafter \@firstoftwo
 \else \expandafter \@secondoftwo
 \fi
}%
\providecommand \natexlab [1]{#1}%
\providecommand \enquote  [1]{``#1''}%
\providecommand \bibnamefont  [1]{#1}%
\providecommand \bibfnamefont [1]{#1}%
\providecommand \citenamefont [1]{#1}%
\providecommand \href@noop [0]{\@secondoftwo}%
\providecommand \href [0]{\begingroup \@sanitize@url \@href}%
\providecommand \@href[1]{\@@startlink{#1}\@@href}%
\providecommand \@@href[1]{\endgroup#1\@@endlink}%
\providecommand \@sanitize@url [0]{\catcode `\\12\catcode `\$12\catcode
  `\&12\catcode `\#12\catcode `\^12\catcode `\_12\catcode `\%12\relax}%
\providecommand \@@startlink[1]{}%
\providecommand \@@endlink[0]{}%
\providecommand \url  [0]{\begingroup\@sanitize@url \@url }%
\providecommand \@url [1]{\endgroup\@href {#1}{\urlprefix }}%
\providecommand \urlprefix  [0]{URL }%
\providecommand \Eprint [0]{\href }%
\providecommand \doibase [0]{http://dx.doi.org/}%
\providecommand \selectlanguage [0]{\@gobble}%
\providecommand \bibinfo  [0]{\@secondoftwo}%
\providecommand \bibfield  [0]{\@secondoftwo}%
\providecommand \translation [1]{[#1]}%
\providecommand \BibitemOpen [0]{}%
\providecommand \bibitemStop [0]{}%
\providecommand \bibitemNoStop [0]{.\EOS\space}%
\providecommand \EOS [0]{\spacefactor3000\relax}%
\providecommand \BibitemShut  [1]{\csname bibitem#1\endcsname}%
\let\auto@bib@innerbib\@empty
\bibitem [{\citenamefont {Anderson}\ and\ \citenamefont
  {May}(1991)}]{Anderson:1991qr}%
  \BibitemOpen
  \bibfield  {author} {\bibinfo {author} {\bibfnamefont {R.~M.}\ \bibnamefont
  {Anderson}}\ and\ \bibinfo {author} {\bibfnamefont {R.~M.}\ \bibnamefont
  {May}},\ }\href@noop {} {\emph {\bibinfo {title} {{Infectious Diseases of
  Humans}}}}\ (\bibinfo  {publisher} {Oxford University Press},\ \bibinfo
  {address} {Oxford},\ \bibinfo {year} {1991})\BibitemShut {NoStop}%
\bibitem [{\citenamefont {Pastor-Satorras}\ and\ \citenamefont
  {Vespignani}(2001)}]{Pastor-Satorras:2001fl}%
  \BibitemOpen
  \bibfield  {author} {\bibinfo {author} {\bibfnamefont {R.}~\bibnamefont
  {Pastor-Satorras}}\ and\ \bibinfo {author} {\bibfnamefont {A.}~\bibnamefont
  {Vespignani}},\ }\href@noop {} {\bibfield  {journal} {\bibinfo  {journal}
  {Phys. Rev. Lett.}\ }\textbf {\bibinfo {volume} {86}},\ \bibinfo {pages}
  {3200} (\bibinfo {year} {2001})}\BibitemShut {NoStop}%
\bibitem [{\citenamefont {Bond}\ \emph {et~al.}(2012)\citenamefont {Bond},
  \citenamefont {Fariss}, \citenamefont {Jones}, \citenamefont {Kramer},
  \citenamefont {Marlow}, \citenamefont {Settle},\ and\ \citenamefont
  {Fowler}}]{Bond:2012xy}%
  \BibitemOpen
  \bibfield  {author} {\bibinfo {author} {\bibfnamefont {R.~M.}\ \bibnamefont
  {Bond}}, \bibinfo {author} {\bibfnamefont {C.~J.}\ \bibnamefont {Fariss}},
  \bibinfo {author} {\bibfnamefont {J.~J.}\ \bibnamefont {Jones}}, \bibinfo
  {author} {\bibfnamefont {A.~D.~I.}\ \bibnamefont {Kramer}}, \bibinfo {author}
  {\bibfnamefont {C.}~\bibnamefont {Marlow}}, \bibinfo {author} {\bibfnamefont
  {J.~E.}\ \bibnamefont {Settle}}, \ and\ \bibinfo {author} {\bibfnamefont
  {J.~H.}\ \bibnamefont {Fowler}},\ }\href
  {http://dx.doi.org/10.1038/nature11421} {\bibfield  {journal} {\bibinfo
  {journal} {Nature}\ }\textbf {\bibinfo {volume} {489}},\ \bibinfo {pages}
  {295} (\bibinfo {year} {2012})}\BibitemShut {NoStop}%
\bibitem [{\citenamefont {Colizza}\ \emph {et~al.}(2007)\citenamefont
  {Colizza}, \citenamefont {Barrat}, \citenamefont {Barthelemy}, \citenamefont
  {Valleron},\ and\ \citenamefont {Vespignani}}]{Colizza:2007aa}%
  \BibitemOpen
  \bibfield  {author} {\bibinfo {author} {\bibfnamefont {V.}~\bibnamefont
  {Colizza}}, \bibinfo {author} {\bibfnamefont {A.}~\bibnamefont {Barrat}},
  \bibinfo {author} {\bibfnamefont {M.}~\bibnamefont {Barthelemy}}, \bibinfo
  {author} {\bibfnamefont {A.-J.}\ \bibnamefont {Valleron}}, \ and\ \bibinfo
  {author} {\bibfnamefont {A.}~\bibnamefont {Vespignani}},\ }\href {\doibase
  10.1371/journal.pmed.0040013} {\bibfield  {journal} {\bibinfo  {journal}
  {PLOS Med.}\ }\textbf {\bibinfo {volume} {4}},\ \bibinfo {pages} {e13}
  (\bibinfo {year} {2007})}\BibitemShut {NoStop}%
\bibitem [{\citenamefont {Vespignani}(2012)}]{Vespignani:2012}%
  \BibitemOpen
  \bibfield  {author} {\bibinfo {author} {\bibfnamefont {A.}~\bibnamefont
  {Vespignani}},\ }\href@noop {} {\bibfield  {journal} {\bibinfo  {journal}
  {Nat. Phys.}\ }\textbf {\bibinfo {volume} {8}},\ \bibinfo {pages} {32}
  (\bibinfo {year} {2012})}\BibitemShut {NoStop}%
\bibitem [{\citenamefont {Pastor-Satorras}\ \emph {et~al.}(2015)\citenamefont
  {Pastor-Satorras}, \citenamefont {Castellano}, \citenamefont {{Van
  Mieghem}},\ and\ \citenamefont {Vespignani}}]{Pastor-Satorras:2015qf}%
  \BibitemOpen
  \bibfield  {author} {\bibinfo {author} {\bibfnamefont {R.}~\bibnamefont
  {Pastor-Satorras}}, \bibinfo {author} {\bibfnamefont {C.}~\bibnamefont
  {Castellano}}, \bibinfo {author} {\bibfnamefont {P.}~\bibnamefont {{Van
  Mieghem}}}, \ and\ \bibinfo {author} {\bibfnamefont {A.}~\bibnamefont
  {Vespignani}},\ }\href {\doibase 10.1103/RevModPhys.87.925} {\bibfield
  {journal} {\bibinfo  {journal} {Rev. Mod. Phys.}\ }\textbf {\bibinfo {volume}
  {87}},\ \bibinfo {pages} {925} (\bibinfo {year} {2015})}\BibitemShut
  {NoStop}%
\bibitem [{\citenamefont {Colizza}\ \emph {et~al.}(2006)\citenamefont
  {Colizza}, \citenamefont {Barrat}, \citenamefont {Barthelemy},\ and\
  \citenamefont {Vespignani}}]{Colizza:2006pp}%
  \BibitemOpen
  \bibfield  {author} {\bibinfo {author} {\bibfnamefont {V.}~\bibnamefont
  {Colizza}}, \bibinfo {author} {\bibfnamefont {A.}~\bibnamefont {Barrat}},
  \bibinfo {author} {\bibfnamefont {M.}~\bibnamefont {Barthelemy}}, \ and\
  \bibinfo {author} {\bibfnamefont {A.}~\bibnamefont {Vespignani}},\
  }\href@noop {} {\bibfield  {journal} {\bibinfo  {journal} {Proc. Natl. Acad.
  Sci. U.S.A.}\ }\textbf {\bibinfo {volume} {103}},\ \bibinfo {pages} {2015}
  (\bibinfo {year} {2006})}\BibitemShut {NoStop}%
\bibitem [{\citenamefont {Balcan}\ \emph {et~al.}(2009)\citenamefont {Balcan},
  \citenamefont {Colizza}, \citenamefont {Gon\c{c}alves}, \citenamefont {Hu},
  \citenamefont {Ramasco},\ and\ \citenamefont {Vespignani}}]{Balcan22122009}%
  \BibitemOpen
  \bibfield  {author} {\bibinfo {author} {\bibfnamefont {D.}~\bibnamefont
  {Balcan}}, \bibinfo {author} {\bibfnamefont {V.}~\bibnamefont {Colizza}},
  \bibinfo {author} {\bibfnamefont {B.}~\bibnamefont {Gon\c{c}alves}}, \bibinfo
  {author} {\bibfnamefont {H.}~\bibnamefont {Hu}}, \bibinfo {author}
  {\bibfnamefont {J.~J.}\ \bibnamefont {Ramasco}}, \ and\ \bibinfo {author}
  {\bibfnamefont {A.}~\bibnamefont {Vespignani}},\ }\href {\doibase
  10.1073/pnas.0906910106} {\bibfield  {journal} {\bibinfo  {journal} {Proc.
  Natl. Acad. Sci. U.S.A.}\ }\textbf {\bibinfo {volume} {106}},\ \bibinfo
  {pages} {21484} (\bibinfo {year} {2009})}\BibitemShut {NoStop}%
\bibitem [{\citenamefont {Tizzoni}\ \emph {et~al.}(2012)\citenamefont
  {Tizzoni}, \citenamefont {Bajardi}, \citenamefont {Poletto}, \citenamefont
  {Ramasco}, \citenamefont {Balcan}, \citenamefont {Gon\c{c}alves},
  \citenamefont {Perra}, \citenamefont {Colizza},\ and\ \citenamefont
  {Vespignani}}]{Tizzoni:2012}%
  \BibitemOpen
  \bibfield  {author} {\bibinfo {author} {\bibfnamefont {M.}~\bibnamefont
  {Tizzoni}}, \bibinfo {author} {\bibfnamefont {P.}~\bibnamefont {Bajardi}},
  \bibinfo {author} {\bibfnamefont {C.}~\bibnamefont {Poletto}}, \bibinfo
  {author} {\bibfnamefont {J.~J.}\ \bibnamefont {Ramasco}}, \bibinfo {author}
  {\bibfnamefont {D.}~\bibnamefont {Balcan}}, \bibinfo {author} {\bibfnamefont
  {B.}~\bibnamefont {Gon\c{c}alves}}, \bibinfo {author} {\bibfnamefont
  {N.}~\bibnamefont {Perra}}, \bibinfo {author} {\bibfnamefont
  {V.}~\bibnamefont {Colizza}}, \ and\ \bibinfo {author} {\bibfnamefont
  {A.}~\bibnamefont {Vespignani}},\ }\href {\doibase 10.1186/1741-7015-10-165}
  {\bibfield  {journal} {\bibinfo  {journal} {BMC Med.}\ }\textbf {\bibinfo
  {volume} {10}},\ \bibinfo {pages} {165} (\bibinfo {year} {2012})}\BibitemShut
  {NoStop}%
\bibitem [{\citenamefont {Streftaris}\ and\ \citenamefont
  {Gibson}(2012)}]{Streftaris:2012}%
  \BibitemOpen
  \bibfield  {author} {\bibinfo {author} {\bibfnamefont {G.}~\bibnamefont
  {Streftaris}}\ and\ \bibinfo {author} {\bibfnamefont {G.~J.}\ \bibnamefont
  {Gibson}},\ }\href {\doibase 10.1093/biostatistics/kxs011} {\bibfield
  {journal} {\bibinfo  {journal} {Biostatistics}\ }\textbf {\bibinfo {volume}
  {13}},\ \bibinfo {pages} {580} (\bibinfo {year} {2012})}\BibitemShut
  {NoStop}%
\bibitem [{\citenamefont {Chowell}\ and\ \citenamefont
  {Nishiura}(2014)}]{Chowell:2014aa}%
  \BibitemOpen
  \bibfield  {author} {\bibinfo {author} {\bibfnamefont {G.}~\bibnamefont
  {Chowell}}\ and\ \bibinfo {author} {\bibfnamefont {H.}~\bibnamefont
  {Nishiura}},\ }\href {\doibase 10.1186/s12916-014-0196-0} {\bibfield
  {journal} {\bibinfo  {journal} {BMC Med.}\ }\textbf {\bibinfo {volume}
  {12}},\ \bibinfo {pages} {196} (\bibinfo {year} {2014})}\BibitemShut
  {NoStop}%
\bibitem [{\citenamefont {Oliveira}\ and\ \citenamefont
  {Barabasi}(2005)}]{Oliveira:2005fk}%
  \BibitemOpen
  \bibfield  {author} {\bibinfo {author} {\bibfnamefont {J.~G.}\ \bibnamefont
  {Oliveira}}\ and\ \bibinfo {author} {\bibfnamefont {A.-L.}\ \bibnamefont
  {Barabasi}},\ }\href@noop {} {\bibfield  {journal} {\bibinfo  {journal}
  {Nature}\ }\textbf {\bibinfo {volume} {437}},\ \bibinfo {pages} {1251}
  (\bibinfo {year} {2005})}\BibitemShut {NoStop}%
\bibitem [{\citenamefont {Gonz{\'a}lez}\ \emph {et~al.}(2008)\citenamefont
  {Gonz{\'a}lez}, \citenamefont {Hidalgo},\ and\ \citenamefont
  {Barab{\'a}si}}]{Gonzalez:2008fk}%
  \BibitemOpen
  \bibfield  {author} {\bibinfo {author} {\bibfnamefont {M.~C.}\ \bibnamefont
  {Gonz{\'a}lez}}, \bibinfo {author} {\bibfnamefont {C.~A.}\ \bibnamefont
  {Hidalgo}}, \ and\ \bibinfo {author} {\bibfnamefont {A.-L.}\ \bibnamefont
  {Barab{\'a}si}},\ }\href@noop {} {\bibfield  {journal} {\bibinfo  {journal}
  {Nature}\ }\textbf {\bibinfo {volume} {453}},\ \bibinfo {pages} {779}
  (\bibinfo {year} {2008})}\BibitemShut {NoStop}%
\bibitem [{\citenamefont {Ben-Jacob}\ \emph {et~al.}(1994)\citenamefont
  {Ben-Jacob}, \citenamefont {Schochet}, \citenamefont {Tenenbaum},
  \citenamefont {Cohen}, \citenamefont {Czir{\'o}k},\ and\ \citenamefont
  {Vicsek}}]{Ben-Jacob:1994}%
  \BibitemOpen
  \bibfield  {author} {\bibinfo {author} {\bibfnamefont {E.}~\bibnamefont
  {Ben-Jacob}}, \bibinfo {author} {\bibfnamefont {O.}~\bibnamefont {Schochet}},
  \bibinfo {author} {\bibfnamefont {A.}~\bibnamefont {Tenenbaum}}, \bibinfo
  {author} {\bibfnamefont {I.}~\bibnamefont {Cohen}}, \bibinfo {author}
  {\bibfnamefont {A.}~\bibnamefont {Czir{\'o}k}}, \ and\ \bibinfo {author}
  {\bibfnamefont {T.}~\bibnamefont {Vicsek}},\ }\href@noop {} {\bibfield
  {journal} {\bibinfo  {journal} {Nature}\ }\textbf {\bibinfo {volume} {368}},\
  \bibinfo {pages} {46} (\bibinfo {year} {1994})}\BibitemShut {NoStop}%
\bibitem [{\citenamefont {Donabedian}(2003)}]{Donabedian:2003}%
  \BibitemOpen
  \bibfield  {author} {\bibinfo {author} {\bibfnamefont {H.}~\bibnamefont
  {Donabedian}},\ }\href@noop {} {\bibfield  {journal} {\bibinfo  {journal} {J.
  Infect.}\ }\textbf {\bibinfo {volume} {46}},\ \bibinfo {pages} {207}
  (\bibinfo {year} {2003})}\BibitemShut {NoStop}%
\bibitem [{\citenamefont {Castellano}\ \emph {et~al.}(2009)\citenamefont
  {Castellano}, \citenamefont {Fortunato},\ and\ \citenamefont
  {Loreto}}]{Castellano:2008}%
  \BibitemOpen
  \bibfield  {author} {\bibinfo {author} {\bibfnamefont {C.}~\bibnamefont
  {Castellano}}, \bibinfo {author} {\bibfnamefont {S.}~\bibnamefont
  {Fortunato}}, \ and\ \bibinfo {author} {\bibfnamefont {V.}~\bibnamefont
  {Loreto}},\ }\href {\doibase 10.1103/RevModPhys.81.591} {\bibfield  {journal}
  {\bibinfo  {journal} {Rev. Mod. Phys.}\ }\textbf {\bibinfo {volume} {81}},\
  \bibinfo {pages} {591} (\bibinfo {year} {2009})}\BibitemShut {NoStop}%
\bibitem [{\citenamefont {Centola}(2010)}]{Centola:2010}%
  \BibitemOpen
  \bibfield  {author} {\bibinfo {author} {\bibfnamefont {D.}~\bibnamefont
  {Centola}},\ }\href {\doibase 10.1126/science.1185231} {\bibfield  {journal}
  {\bibinfo  {journal} {Science}\ }\textbf {\bibinfo {volume} {329}},\ \bibinfo
  {pages} {1194} (\bibinfo {year} {2010})}\BibitemShut {NoStop}%
\bibitem [{\citenamefont {Hodas}\ and\ \citenamefont
  {Lerman}(2014)}]{Hodas:2014}%
  \BibitemOpen
  \bibfield  {author} {\bibinfo {author} {\bibfnamefont {N.~O.}\ \bibnamefont
  {Hodas}}\ and\ \bibinfo {author} {\bibfnamefont {K.}~\bibnamefont {Lerman}},\
  }\href@noop {} {\bibfield  {journal} {\bibinfo  {journal} {Sci. Rep.}\
  }\textbf {\bibinfo {volume} {4}},\ \bibinfo {pages} {4343} (\bibinfo {year}
  {2014})}\BibitemShut {NoStop}%
\bibitem [{\citenamefont {Choi}\ \emph
  {et~al.}(2017{\natexlab{a}})\citenamefont {Choi}, \citenamefont {Lee},\ and\
  \citenamefont {Kahng}}]{Choi:2017a}%
  \BibitemOpen
  \bibfield  {author} {\bibinfo {author} {\bibfnamefont {W.}~\bibnamefont
  {Choi}}, \bibinfo {author} {\bibfnamefont {D.}~\bibnamefont {Lee}}, \ and\
  \bibinfo {author} {\bibfnamefont {B.}~\bibnamefont {Kahng}},\ }\href
  {\doibase 10.1103/PhysRevE.95.022304} {\bibfield  {journal} {\bibinfo
  {journal} {Phys. Rev. E}\ }\textbf {\bibinfo {volume} {95}},\ \bibinfo
  {pages} {022304} (\bibinfo {year} {2017}{\natexlab{a}})}\BibitemShut
  {NoStop}%
\bibitem [{\citenamefont {Choi}\ \emph
  {et~al.}(2017{\natexlab{b}})\citenamefont {Choi}, \citenamefont {Lee},\ and\
  \citenamefont {Kahng}}]{Choi:2017b}%
  \BibitemOpen
  \bibfield  {author} {\bibinfo {author} {\bibfnamefont {W.}~\bibnamefont
  {Choi}}, \bibinfo {author} {\bibfnamefont {D.}~\bibnamefont {Lee}}, \ and\
  \bibinfo {author} {\bibfnamefont {B.}~\bibnamefont {Kahng}},\ }\href
  {\doibase 10.1103/PhysRevE.95.062115} {\bibfield  {journal} {\bibinfo
  {journal} {Phys. Rev. E}\ }\textbf {\bibinfo {volume} {95}},\ \bibinfo
  {pages} {062115} (\bibinfo {year} {2017}{\natexlab{b}})}\BibitemShut
  {NoStop}%
\bibitem [{\citenamefont {{Van Mieghem}}\ and\ \citenamefont {{van de
  Bovenkamp}}(2013)}]{Van-Mieghem:2013db}%
  \BibitemOpen
  \bibfield  {author} {\bibinfo {author} {\bibfnamefont {P.}~\bibnamefont {{Van
  Mieghem}}}\ and\ \bibinfo {author} {\bibfnamefont {R.}~\bibnamefont {{van de
  Bovenkamp}}},\ }\href {\doibase 10.1103/PhysRevLett.110.108701} {\bibfield
  {journal} {\bibinfo  {journal} {Phys. Rev. Lett.}\ }\textbf {\bibinfo
  {volume} {110}},\ \bibinfo {pages} {108701} (\bibinfo {year}
  {2013})}\BibitemShut {NoStop}%
\bibitem [{\citenamefont {Kiss}\ \emph {et~al.}(2015)\citenamefont {Kiss},
  \citenamefont {R{\"o}st},\ and\ \citenamefont {Vizi}}]{Kiss:2015}%
  \BibitemOpen
  \bibfield  {author} {\bibinfo {author} {\bibfnamefont {I.~Z.}\ \bibnamefont
  {Kiss}}, \bibinfo {author} {\bibfnamefont {G.}~\bibnamefont {R{\"o}st}}, \
  and\ \bibinfo {author} {\bibfnamefont {Z.}~\bibnamefont {Vizi}},\ }\href
  {\doibase 10.1103/PhysRevLett.115.078701} {\bibfield  {journal} {\bibinfo
  {journal} {Phys. Rev. Lett.}\ }\textbf {\bibinfo {volume} {115}},\ \bibinfo
  {pages} {078701} (\bibinfo {year} {2015})}\BibitemShut {NoStop}%
\bibitem [{\citenamefont {Starnini}\ \emph {et~al.}(2017)\citenamefont
  {Starnini}, \citenamefont {Gleeson},\ and\ \citenamefont
  {Bogu{\~n}{\'a}}}]{Starnini:2017}%
  \BibitemOpen
  \bibfield  {author} {\bibinfo {author} {\bibfnamefont {M.}~\bibnamefont
  {Starnini}}, \bibinfo {author} {\bibfnamefont {J.~P.}\ \bibnamefont
  {Gleeson}}, \ and\ \bibinfo {author} {\bibfnamefont {M.}~\bibnamefont
  {Bogu{\~n}{\'a}}},\ }\href {\doibase 10.1103/PhysRevLett.118.128301}
  {\bibfield  {journal} {\bibinfo  {journal} {Phys. Rev. Lett.}\ }\textbf
  {\bibinfo {volume} {118}},\ \bibinfo {pages} {128301} (\bibinfo {year}
  {2017})}\BibitemShut {NoStop}%
\bibitem [{\citenamefont {Dodds}\ and\ \citenamefont
  {Watts}(2004)}]{Dodds:2004}%
  \BibitemOpen
  \bibfield  {author} {\bibinfo {author} {\bibfnamefont {P.~S.}\ \bibnamefont
  {Dodds}}\ and\ \bibinfo {author} {\bibfnamefont {D.~J.}\ \bibnamefont
  {Watts}},\ }\href {\doibase 10.1103/PhysRevLett.92.218701} {\bibfield
  {journal} {\bibinfo  {journal} {Phys. Rev. Lett.}\ }\textbf {\bibinfo
  {volume} {92}},\ \bibinfo {pages} {218701} (\bibinfo {year}
  {2004})}\BibitemShut {NoStop}%
\bibitem [{\citenamefont {L{\"u}}\ \emph {et~al.}(2011)\citenamefont {L{\"u}},
  \citenamefont {Chen},\ and\ \citenamefont {Zhou}}]{Lu:2011}%
  \BibitemOpen
  \bibfield  {author} {\bibinfo {author} {\bibfnamefont {L.}~\bibnamefont
  {L{\"u}}}, \bibinfo {author} {\bibfnamefont {D.-B.}\ \bibnamefont {Chen}}, \
  and\ \bibinfo {author} {\bibfnamefont {T.}~\bibnamefont {Zhou}},\ }\href
  {http://stacks.iop.org/1367-2630/13/i=12/a=123005} {\bibfield  {journal}
  {\bibinfo  {journal} {New J. Phys.}\ }\textbf {\bibinfo {volume} {13}},\
  \bibinfo {pages} {123005} (\bibinfo {year} {2011})}\BibitemShut {NoStop}%
\bibitem [{\citenamefont {Liu}\ \emph {et~al.}(1987)\citenamefont {Liu},
  \citenamefont {Hethcote},\ and\ \citenamefont {Levin}}]{Liu:1987}%
  \BibitemOpen
  \bibfield  {author} {\bibinfo {author} {\bibfnamefont {W.}~\bibnamefont
  {Liu}}, \bibinfo {author} {\bibfnamefont {H.~W.}\ \bibnamefont {Hethcote}}, \
  and\ \bibinfo {author} {\bibfnamefont {S.~A.}\ \bibnamefont {Levin}},\ }\href
  {\doibase 10.1007/BF00277162} {\bibfield  {journal} {\bibinfo  {journal} {J.
  Math. Biol.}\ }\textbf {\bibinfo {volume} {25}},\ \bibinfo {pages} {359}
  (\bibinfo {year} {1987})}\BibitemShut {NoStop}%
\bibitem [{\citenamefont {Bogu{\~n}{\'a}}\ \emph {et~al.}(2014)\citenamefont
  {Bogu{\~n}{\'a}}, \citenamefont {Lafuerza}, \citenamefont {Toral},\ and\
  \citenamefont {Serrano}}]{Boguna:2014}%
  \BibitemOpen
  \bibfield  {author} {\bibinfo {author} {\bibfnamefont {M.}~\bibnamefont
  {Bogu{\~n}{\'a}}}, \bibinfo {author} {\bibfnamefont {L.~F.}\ \bibnamefont
  {Lafuerza}}, \bibinfo {author} {\bibfnamefont {R.}~\bibnamefont {Toral}}, \
  and\ \bibinfo {author} {\bibfnamefont {M.~{\'A}.}\ \bibnamefont {Serrano}},\
  }\href {\doibase 10.1103/PhysRevE.90.042108} {\bibfield  {journal} {\bibinfo
  {journal} {Phys. Rev. E}\ }\textbf {\bibinfo {volume} {90}},\ \bibinfo
  {pages} {042108} (\bibinfo {year} {2014})}\BibitemShut {NoStop}%
\bibitem [{\citenamefont {P{\'e}rez-Reche}\ \emph {et~al.}(2011)\citenamefont
  {P{\'e}rez-Reche}, \citenamefont {Ludlam}, \citenamefont {Taraskin},\ and\
  \citenamefont {Gilligan}}]{Perez-Reche:2011}%
  \BibitemOpen
  \bibfield  {author} {\bibinfo {author} {\bibfnamefont {F.~J.}\ \bibnamefont
  {P{\'e}rez-Reche}}, \bibinfo {author} {\bibfnamefont {J.~J.}\ \bibnamefont
  {Ludlam}}, \bibinfo {author} {\bibfnamefont {S.~N.}\ \bibnamefont
  {Taraskin}}, \ and\ \bibinfo {author} {\bibfnamefont {C.~A.}\ \bibnamefont
  {Gilligan}},\ }\href {\doibase 10.1103/PhysRevLett.106.218701} {\bibfield
  {journal} {\bibinfo  {journal} {Phys. Rev. Lett.}\ }\textbf {\bibinfo
  {volume} {106}},\ \bibinfo {pages} {218701} (\bibinfo {year}
  {2011})}\BibitemShut {NoStop}%
\bibitem [{\citenamefont {G{\'o}mez-Garde{\~n}es}\ \emph
  {et~al.}(2016)\citenamefont {G{\'o}mez-Garde{\~n}es}, \citenamefont {Lotero},
  \citenamefont {Taraskin},\ and\ \citenamefont
  {P{\'e}rez-Reche}}]{Gomez-Gardenes:2016}%
  \BibitemOpen
  \bibfield  {author} {\bibinfo {author} {\bibfnamefont {J.}~\bibnamefont
  {G{\'o}mez-Garde{\~n}es}}, \bibinfo {author} {\bibfnamefont {L.}~\bibnamefont
  {Lotero}}, \bibinfo {author} {\bibfnamefont {S.~N.}\ \bibnamefont
  {Taraskin}}, \ and\ \bibinfo {author} {\bibfnamefont {F.~J.}\ \bibnamefont
  {P{\'e}rez-Reche}},\ }\href@noop {} {\bibfield  {journal} {\bibinfo
  {journal} {Sci. Rep.}\ }\textbf {\bibinfo {volume} {6}},\ \bibinfo {pages}
  {19767} (\bibinfo {year} {2016})}\BibitemShut {NoStop}%
\bibitem [{\citenamefont {Watts}(2002)}]{Watts:2002}%
  \BibitemOpen
  \bibfield  {author} {\bibinfo {author} {\bibfnamefont {D.~J.}\ \bibnamefont
  {Watts}},\ }\href {\doibase 10.1073/pnas.082090499} {\bibfield  {journal}
  {\bibinfo  {journal} {Proc. Natl. Acad. Sci. U.S.A.}\ }\textbf {\bibinfo
  {volume} {99}},\ \bibinfo {pages} {5766} (\bibinfo {year}
  {2002})}\BibitemShut {NoStop}%
\bibitem [{\citenamefont {Centola}\ \emph {et~al.}(2007)\citenamefont
  {Centola}, \citenamefont {Egu{\'\i}luz},\ and\ \citenamefont
  {Macy}}]{Centola:2007}%
  \BibitemOpen
  \bibfield  {author} {\bibinfo {author} {\bibfnamefont {D.}~\bibnamefont
  {Centola}}, \bibinfo {author} {\bibfnamefont {V.~M.}\ \bibnamefont
  {Egu{\'\i}luz}}, \ and\ \bibinfo {author} {\bibfnamefont {M.~W.}\
  \bibnamefont {Macy}},\ }\href {\doibase 10.1016/j.physa.2006.06.018}
  {\bibfield  {journal} {\bibinfo  {journal} {Physica A}\ }\textbf {\bibinfo
  {volume} {374}},\ \bibinfo {pages} {449} (\bibinfo {year}
  {2007})}\BibitemShut {NoStop}%
\bibitem [{\citenamefont {Granell}\ \emph {et~al.}(2013)\citenamefont
  {Granell}, \citenamefont {G\'omez},\ and\ \citenamefont
  {Arenas}}]{Granell:2013}%
  \BibitemOpen
  \bibfield  {author} {\bibinfo {author} {\bibfnamefont {C.}~\bibnamefont
  {Granell}}, \bibinfo {author} {\bibfnamefont {S.}~\bibnamefont {G\'omez}}, \
  and\ \bibinfo {author} {\bibfnamefont {A.}~\bibnamefont {Arenas}},\ }\href
  {\doibase 10.1103/PhysRevLett.111.128701} {\bibfield  {journal} {\bibinfo
  {journal} {Phys. Rev. Lett.}\ }\textbf {\bibinfo {volume} {111}},\ \bibinfo
  {pages} {128701} (\bibinfo {year} {2013})}\BibitemShut {NoStop}%
\bibitem [{\citenamefont {Fu}\ \emph {et~al.}(2017)\citenamefont {Fu},
  \citenamefont {Christakis},\ and\ \citenamefont {Fowler}}]{Fu:2017}%
  \BibitemOpen
  \bibfield  {author} {\bibinfo {author} {\bibfnamefont {F.}~\bibnamefont
  {Fu}}, \bibinfo {author} {\bibfnamefont {N.~A.}\ \bibnamefont {Christakis}},
  \ and\ \bibinfo {author} {\bibfnamefont {J.~H.}\ \bibnamefont {Fowler}},\
  }\href {http://dx.doi.org/10.1038/srep43634} {\bibfield  {journal} {\bibinfo
  {journal} {Sci. Rep.}\ }\textbf {\bibinfo {volume} {7}},\ \bibinfo {pages}
  {43634} (\bibinfo {year} {2017})}\BibitemShut {NoStop}%
\bibitem [{\citenamefont {Christakis}\ and\ \citenamefont
  {Fowler}(2007)}]{Christakis:2007}%
  \BibitemOpen
  \bibfield  {author} {\bibinfo {author} {\bibfnamefont {N.~A.}\ \bibnamefont
  {Christakis}}\ and\ \bibinfo {author} {\bibfnamefont {J.~H.}\ \bibnamefont
  {Fowler}},\ }\href {\doibase 10.1056/NEJMsa066082} {\bibfield  {journal}
  {\bibinfo  {journal} {N. Engl. J. Med.}\ }\textbf {\bibinfo {volume} {357}},\
  \bibinfo {pages} {370} (\bibinfo {year} {2007})}\BibitemShut {NoStop}%
\bibitem [{\citenamefont {Christakis}\ and\ \citenamefont
  {Fowler}(2008)}]{Christakis:2008}%
  \BibitemOpen
  \bibfield  {author} {\bibinfo {author} {\bibfnamefont {N.~A.}\ \bibnamefont
  {Christakis}}\ and\ \bibinfo {author} {\bibfnamefont {J.~H.}\ \bibnamefont
  {Fowler}},\ }\href {\doibase 10.1056/NEJMsa0706154} {\bibfield  {journal}
  {\bibinfo  {journal} {N. Engl. J. Med.}\ }\textbf {\bibinfo {volume} {358}},\
  \bibinfo {pages} {2249} (\bibinfo {year} {2008})}\BibitemShut {NoStop}%
\bibitem [{\citenamefont {Cox}(1970)}]{cox:1970}%
  \BibitemOpen
  \bibfield  {author} {\bibinfo {author} {\bibfnamefont {D.}~\bibnamefont
  {Cox}},\ }\href@noop {} {\emph {\bibinfo {title} {{Renewal theory}}}}\
  (\bibinfo  {publisher} {Methuen \& Co.},\ \bibinfo {address} {London},\
  \bibinfo {year} {1970})\BibitemShut {NoStop}%
\bibitem [{\citenamefont {Liu}\ and\ \citenamefont {{Van
  Mieghem}}(2018)}]{Liu:2018}%
  \BibitemOpen
  \bibfield  {author} {\bibinfo {author} {\bibfnamefont {Q.}~\bibnamefont
  {Liu}}\ and\ \bibinfo {author} {\bibfnamefont {P.}~\bibnamefont {{Van
  Mieghem}}},\ }\href {\doibase 10.1103/PhysRevE.97.022309} {\bibfield
  {journal} {\bibinfo  {journal} {Phys. Rev. E}\ }\textbf {\bibinfo {volume}
  {97}},\ \bibinfo {pages} {022309} (\bibinfo {year} {2018})}\BibitemShut
  {NoStop}%
\bibitem [{\citenamefont {Min}\ and\ \citenamefont
  {San~Miguel}(2018)}]{Min:2018}%
  \BibitemOpen
  \bibfield  {author} {\bibinfo {author} {\bibfnamefont {B.}~\bibnamefont
  {Min}}\ and\ \bibinfo {author} {\bibfnamefont {M.}~\bibnamefont
  {San~Miguel}},\ }\href {\doibase 10.1038/s41598-018-28615-3} {\bibfield
  {journal} {\bibinfo  {journal} {Scientific Reports}\ }\textbf {\bibinfo
  {volume} {8}},\ \bibinfo {pages} {10422} (\bibinfo {year}
  {2018})}\BibitemShut {NoStop}%
\bibitem [{\citenamefont {Mata}\ \emph {et~al.}(2015)\citenamefont {Mata},
  \citenamefont {Bogu{\~n}{\'a}}, \citenamefont {Castellano},\ and\
  \citenamefont {Pastor-Satorras}}]{Mata:2015}%
  \BibitemOpen
  \bibfield  {author} {\bibinfo {author} {\bibfnamefont {A.~S.}\ \bibnamefont
  {Mata}}, \bibinfo {author} {\bibfnamefont {M.}~\bibnamefont
  {Bogu{\~n}{\'a}}}, \bibinfo {author} {\bibfnamefont {C.}~\bibnamefont
  {Castellano}}, \ and\ \bibinfo {author} {\bibfnamefont {R.}~\bibnamefont
  {Pastor-Satorras}},\ }\href {\doibase 10.1103/PhysRevE.91.052117} {\bibfield
  {journal} {\bibinfo  {journal} {Phys. Rev. E}\ }\textbf {\bibinfo {volume}
  {91}},\ \bibinfo {pages} {052117} (\bibinfo {year} {2015})}\BibitemShut
  {NoStop}%
\bibitem [{\citenamefont {Chen}\ \emph {et~al.}(2018)\citenamefont {Chen},
  \citenamefont {Wang}, \citenamefont {Tang}, \citenamefont {Cai},
  \citenamefont {Stanley},\ and\ \citenamefont {Braunstein}}]{Chen:2018}%
  \BibitemOpen
  \bibfield  {author} {\bibinfo {author} {\bibfnamefont {X.}~\bibnamefont
  {Chen}}, \bibinfo {author} {\bibfnamefont {R.}~\bibnamefont {Wang}}, \bibinfo
  {author} {\bibfnamefont {M.}~\bibnamefont {Tang}}, \bibinfo {author}
  {\bibfnamefont {S.}~\bibnamefont {Cai}}, \bibinfo {author} {\bibfnamefont
  {H.~E.}\ \bibnamefont {Stanley}}, \ and\ \bibinfo {author} {\bibfnamefont
  {L.~A.}\ \bibnamefont {Braunstein}},\ }\href
  {http://stacks.iop.org/1367-2630/20/i=1/a=013007} {\bibfield  {journal}
  {\bibinfo  {journal} {New J. Phys.}\ }\textbf {\bibinfo {volume} {20}},\
  \bibinfo {pages} {013007} (\bibinfo {year} {2018})}\BibitemShut {NoStop}%
\bibitem [{\citenamefont {Cai}\ \emph {et~al.}(2015)\citenamefont {Cai},
  \citenamefont {Chen}, \citenamefont {Ghanbarnejad},\ and\ \citenamefont
  {Grassberger}}]{Cai:2015}%
  \BibitemOpen
  \bibfield  {author} {\bibinfo {author} {\bibfnamefont {W.}~\bibnamefont
  {Cai}}, \bibinfo {author} {\bibfnamefont {L.}~\bibnamefont {Chen}}, \bibinfo
  {author} {\bibfnamefont {F.}~\bibnamefont {Ghanbarnejad}}, \ and\ \bibinfo
  {author} {\bibfnamefont {P.}~\bibnamefont {Grassberger}},\ }\href@noop {}
  {\bibfield  {journal} {\bibinfo  {journal} {Nat. Phys.}\ }\textbf {\bibinfo
  {volume} {11}},\ \bibinfo {pages} {936} (\bibinfo {year} {2015})}\BibitemShut
  {NoStop}%
\bibitem [{\citenamefont {{Colomer-de-Sim\'on}}\ and\ \citenamefont
  {Bogu\~n\'a}(2014)}]{Colomer:2014}%
  \BibitemOpen
  \bibfield  {author} {\bibinfo {author} {\bibfnamefont {P.}~\bibnamefont
  {{Colomer-de-Sim\'on}}}\ and\ \bibinfo {author} {\bibfnamefont
  {M.}~\bibnamefont {Bogu\~n\'a}},\ }\href {\doibase 10.1103/PhysRevX.4.041020}
  {\bibfield  {journal} {\bibinfo  {journal} {Phys. Rev. X}\ }\textbf {\bibinfo
  {volume} {4}},\ \bibinfo {pages} {041020} (\bibinfo {year}
  {2014})}\BibitemShut {NoStop}%
\bibitem [{\citenamefont {Allard}\ \emph {et~al.}(2017)\citenamefont {Allard},
  \citenamefont {Althouse}, \citenamefont {Scarpino},\ and\ \citenamefont
  {H{\'e}bert-Dufresne}}]{Allard:2017}%
  \BibitemOpen
  \bibfield  {author} {\bibinfo {author} {\bibfnamefont {A.}~\bibnamefont
  {Allard}}, \bibinfo {author} {\bibfnamefont {B.~M.}\ \bibnamefont
  {Althouse}}, \bibinfo {author} {\bibfnamefont {S.~V.}\ \bibnamefont
  {Scarpino}}, \ and\ \bibinfo {author} {\bibfnamefont {L.}~\bibnamefont
  {H{\'e}bert-Dufresne}},\ }\href {\doibase 10.1073/pnas.1703073114} {\bibfield
   {journal} {\bibinfo  {journal} {Proc. Natl. Acad. Sci. U.S.A.}\ }\textbf
  {\bibinfo {volume} {114}},\ \bibinfo {pages} {8969} (\bibinfo {year}
  {2017})}\BibitemShut {NoStop}%
\bibitem [{\citenamefont {Liu}\ \emph {et~al.}(2017)\citenamefont {Liu},
  \citenamefont {Wang}, \citenamefont {Tang}, \citenamefont {Zhou},\ and\
  \citenamefont {Lai}}]{Liu:2017}%
  \BibitemOpen
  \bibfield  {author} {\bibinfo {author} {\bibfnamefont {Q.-H.}\ \bibnamefont
  {Liu}}, \bibinfo {author} {\bibfnamefont {W.}~\bibnamefont {Wang}}, \bibinfo
  {author} {\bibfnamefont {M.}~\bibnamefont {Tang}}, \bibinfo {author}
  {\bibfnamefont {T.}~\bibnamefont {Zhou}}, \ and\ \bibinfo {author}
  {\bibfnamefont {Y.-C.}\ \bibnamefont {Lai}},\ }\href {\doibase
  10.1103/PhysRevE.95.042320} {\bibfield  {journal} {\bibinfo  {journal} {Phys.
  Rev. E}\ }\textbf {\bibinfo {volume} {95}},\ \bibinfo {pages} {042320}
  (\bibinfo {year} {2017})}\BibitemShut {NoStop}%
\bibitem [{\citenamefont {O'Sullivan}\ \emph {et~al.}(2015)\citenamefont
  {O'Sullivan}, \citenamefont {O'Keeffe}, \citenamefont {Fennell},\ and\
  \citenamefont {Gleeson}}]{OSullivan:2015}%
  \BibitemOpen
  \bibfield  {author} {\bibinfo {author} {\bibfnamefont {D.~J.}\ \bibnamefont
  {O'Sullivan}}, \bibinfo {author} {\bibfnamefont {G.}~\bibnamefont
  {O'Keeffe}}, \bibinfo {author} {\bibfnamefont {P.}~\bibnamefont {Fennell}}, \
  and\ \bibinfo {author} {\bibfnamefont {J.}~\bibnamefont {Gleeson}},\ }\href
  {\doibase 10.3389/fphy.2015.00071} {\bibfield  {journal} {\bibinfo  {journal}
  {Front. Phys.}\ }\textbf {\bibinfo {volume} {3}},\ \bibinfo {pages} {71}
  (\bibinfo {year} {2015})}\BibitemShut {NoStop}%
\bibitem [{Note1()}]{Note1}%
  \BibitemOpen
  \bibinfo {note} {Notice, however, that the terms $\chi (t)$ and $\Omega (t)$
  are stochastic processes that play an important role in the global
  dynamics.}\BibitemShut {Stop}%
\end{thebibliography}
\end{document}


\title{Supplemental Material ``Memory-induced complex contagion in epidemic spreading''}
\author{Xavier R. Hoffmann}
\email{xrhoffmann@gmail.com}
\affiliation{Departament de F\'isica de la Mat\`eria Condensada, Universitat de Barcelona, Spain}
\affiliation{Universitat de Barcelona Institute of Complex Systems (UBICS), Spain}
\author{Mari\'an Bogu\~n\'a}
\email{marian.boguna@ub.edu}
\affiliation{Departament de F\'isica de la Mat\`eria Condensada, Universitat de Barcelona, Spain}
\affiliation{Universitat de Barcelona Institute of Complex Systems (UBICS), Spain}

\maketitle

\section{Analytics: Short-term memory}
At time~$t$, node~$i$ is described by its state~$n_i(t)$ and its viral load~$\kappa_i(t)$. The former is a discrete variable that can take two values, ${n_i(t)=1}$ if its infected, and ${n_i(t)=0}$ if its healthy (susceptible or dormant). The viral load, on the other hand, is a continuous variable with ${\kappa_i(t)\geq0, \forall t}$. A relevant nonindependent variable is the number of $i$'s neighbors that are infected at time~$t$, ${z_i(t)=\sum_ja_{ij}n_j(t)}$, with $a_{ij}$ the elements of the adjacency matrix. In particular, this variable aids in distinguishing healthy susceptible nodes, ${(1-n_i(t))(1-\delta_{z_i(t)}^0)}$, from healthy dormant nodes, ${(1-n_i(t))\delta_{z_i(t)}^0}$, with $\delta_m^\ell$ the Kr\"onecker function ($\delta_m^\ell=1$ if $m=\ell$, $\delta_m^\ell=0$ if $m\neq\ell$).

The evolution of these variables are governed by a series of microscopic, dichotomous, stochastic processes:
\begin{itemize}
\item Infected node~$i$ recovers, $\xi_i=1$, or remains infected, $\xi_i=0$, given by the instantaneous hazard rate of node~$i$'s recovery, $\eta$. At $\mathcal{O}(\dt)$, the corresponding probabilities are 
\begin{equation}
\xi_i=\left\{\begin{array}{lcl}1&&\eta\text{d}t\\0&&1-\eta\text{d}t\end{array}\right.\ .
\end{equation}
\item Susceptible node~$i$ becomes infected, $\pi_i=1$, or remains susceptible, $\pi_i=0$, given by the instantaneous hazard rate of node~$i$'s infection, $\omega_i(t)=\upsilon\alpha\mu^\alpha z_i(t)\kappa_i^{\alpha-1}(t)$. At $\mathcal{O}(\dt)$, the corresponding probabilities are 
\begin{equation}
\pi_i=\left\{\begin{array}{lcl}1&&\omega_i(t)\text{d}t\\0&&1-\omega_i(t)\text{d}t\end{array}\right.\ .
\end{equation}
\item Susceptible node~$i$ becomes dormant, $\chi_i=1$, or remains susceptible, $\chi_i=0$. This transition occurs if all of node~$i$'s infected neighbors recover. At $\mathcal{O}(\dt)$, this reduces to node~$i$ having a single infected neighbor that recovers, thus
\begin{equation}
\chi_i=\left\{\begin{array}{lcl}1&&\delta_{z_i(t)}^1\eta\text{d}t\\0&&1-\delta_{z_i(t)}^1\eta\text{d}t\end{array}\right.\ .
\end{equation}
\end{itemize}

The state of node~$i$ only changes with the transitions infected to healthy and susceptible to infected. When node~$i$ transitions between susceptible and dormant, the state remains unaltered. Then the equation for the state of node~$i$ at time ${t+\text{d}t}$ is
\begin{equation}
n_i(t+\text{d}t)=n_i(t)(1-\xi_i)+(1-n_i(t))(1-\delta_{z_i(t)}^0)\pi_i\ ,
\end{equation}
where the first term corresponds the node~$i$ being infected and not recovering, while the second term corresponds to node~$i$ being susceptible and becoming infected. 

While node~$i$ is susceptible, its viral load increases proportionally to the number of infected neighbors. Moreover, when node~$i$ transitions from infected to healthy, its viral load is erased instantly. Additionally, with short-term memory, $\zeta=0$, node~$i$'s viral load is erased instantly when it transitions from susceptible to dormant. Thus the equation for the viral load of node~$i$ at time ${t+\text{d}t}$ is
\begin{align}
\kappa_i(t+\text{d}t)={ }&\kappa_i(t)-\kappa_i(t)n_i(t)\xi_i+(1-n_i(t))(1-\delta_{z_i(t)}^0)(1-\pi_i)(1-\chi_i)\upsilon z_i(t)\text{d}t\nonumber\\
&-\kappa_i(t)(1-n_i(t))(1-\delta_{z_i(t)}^0)(1-\pi_i)\chi_i\ .
\end{align}
The first term corresponds to the previously amassed viral load, the second term describes the event where infected node~$i$ recovers and erases its viral load, the third term corresponds to susceptible node~$i$ remaining susceptible (neither recovering nor becoming dormant) and accumulating additional viral load from its $z_i(t)$ infected neighbors (from each at rate~$\upsilon$), and the fourth term describes the event where susceptible node~$i$ becomes dormant and instantly erases its viral load.

In order to obtain the dynamic equations, we first compute the expectation value conditioned on time~$t$, which only affects the stochastic variables
\begin{align}
\text{E}[n_i(t+\text{d}t)|t]&=n_i(t)\text{E}[1-\xi_i]+(1-n_i(t))(1-\delta_{z_i(t)}^0)\text{E}[\pi_i]\\&=n_i(t)(1-\eta\text{d}t)+(1-n_i(t))(1-\delta_{z_i(t)}^0)\omega_i(t)\text{d}t\\&=n_i(t)-n_i(t)\eta\text{d}t+\upsilon\alpha\mu^\alpha(1-n_i(t))z_i(t)\kappa_i^{\alpha-1}(t)\text{d}t\ .
\end{align}
Note that the delta term cancels since ${\delta_{z_i(t)}^0z_i(t)=0}$. Taking the ensemble average we find
\begin{equation}
\langle n_i(t+\text{d}t)\rangle=\langle \text{E}[n_i(t+\text{d}t)|t]\rangle =\langle n_i(t)\ran-\eta\lan n_i(t)\ran \dt+\upsilon\alpha\mu^\alpha\lan(1-n_i(t))z_i(t)\kappa_i^{\alpha-1}(t) \ran \dt
\end{equation}
from where we compute
\begin{equation}
\frac{\text{d}\lan n_i(t)\ran}{\dt}=\frac{\langle n_i(t+\text{d}t)\rangle-\langle n_i(t)\ran}{\dt}=-\eta\lan n_i(t)\ran+\upsilon\alpha\mu^\alpha\lan(1-n_i(t))z_i(t)\kappa_i^{\alpha-1}(t) \ran\ .\label{eq:dndt0}
\end{equation}
Following the same procedure for the viral load, we obtain
\begin{equation}
\frac{\text{d}\lan\kappa_i(t)\ran}{\dt}=-\eta\lan\kappa_i(t)n_i(t)\ran-\eta\lan\kappa_i(t)(1-n_i(t))\delta_{z_i(t)}^1\ran+\upsilon\lan(1-n_i(t))z_i(t)\ran\ ,\label{eq:dkdt0}
\end{equation}
and, in addition, we compute the dynamic equation for $\kappa_i^\gamma(t)$, for an arbitrary value of $\gamma>0$,
\begin{equation}
\frac{\text{d}\lan\kappa^\gamma_i(t)\ran}{\dt}=-\eta\lan\kappa^\gamma_i(t)n_i(t)\ran-\eta\lan\kappa_i^\gamma(t)(1-n_i(t))\delta_{z_i(t)}^1\ran+\gamma\upsilon\lan(1-n_i(t))z_i(t)\kappa^{\gamma-1}_i(t)\ran\ .\label{eq:dkgdt0}
\end{equation}
See section~\ref{subsec:details0} for a detailed derivation of Eqs.~\eqref{eq:dkdt0} and~\eqref{eq:dkgdt0}.

\subsection{Late-time limit}
Taking the late-time limit in Eqs.~\eqref{eq:dndt0}, \eqref{eq:dkdt0} and~\eqref{eq:dkgdt0}, and dropping the dependence with~$t$ we have
\begin{align}
0&=-\eta\lan n_i\ran+\upsilon\alpha\mu^\alpha\lan(1-n_i)z_i\kappa^{\alpha-1}_i\ran\label{eq:zerossn}\\
0&=-\eta\lan \kappa_in_i\ran-\eta\lan\kappa_i(1-n_i)\delta_{z_i}^1\ran+\upsilon\lan(1-n_i)z_i\ran\label{eq:zerossk}\\
0&=-\eta\lan \kappa^{\gamma}_in_i\ran-\eta\lan\kappa^{\gamma}_i(1-n_i)\delta_{z_i}^1\ran+\gamma\upsilon\lan(1-n_i)z_i\kappa^{\gamma-1}_i\ran\label{eq:zerosskb}\ .
\end{align}
Setting $\gamma=\alpha$ in Eq.~\eqref{eq:zerosskb} and combining with Eq.~\eqref{eq:zerossn} yields a pair of equations
\begin{align}
0&=-\lan \kappa^{\alpha}_in_i\ran-\lan\kappa^{\alpha}_i(1-n_i)\delta_{z_i}^1\ran+\mu^{-\alpha}\lan n_i\ran\label{eq:red1}\\
0&=-\lan \kappa_in_i\ran-\lan\kappa_i(1-n_i)\delta_{z_i}^1\ran+\upsilon\eta^{-1}\lan(1-n_i)z_i\ran\ .\label{eq:red2}
\end{align}
The generic term ${\lan \kappa^\gamma_in_i\ran}$ can be expanded as
\begin{align}
\lan \kappa^\gamma_in_i\ran&=\lan \kappa^\gamma_in_i|n_i=1\ran \text{Pr}(n_i=1)+\lan \kappa^\gamma_in_i|n_i=0\ran \text{Pr}(n_i=0)\\
&=\lan \kappa^\gamma_i|n_i=1\ran\lan n_i\ran + 0\times(1-\lan n_i\ran)\ .
\end{align}
Similarly, we find ${\lan \kappa^\gamma_i(1-n_i)\delta_{z_i}^1\ran=\lan \kappa^\gamma_i|X_i=1\ran\text{Pr}(X_i=1)}$, with ${X_i=(n_i=0\cap\delta_{z_i}^1=1)}$, i.e., node~$i$ being susceptible and having a single infected neighbor. Substituting for $\gamma=1$ and $\gamma=\alpha$ in Eqs.~\eqref{eq:red1} and~\eqref{eq:red2}, and combining both equations yields
\begin{equation}
-\mu^{-\alpha}\lan n_i\ran+A\upsilon\eta^{-1}\lan(1-n_i)z_i\ran-B\,\text{Pr}(X_i=1)=0\ ,\label{eq:singleeq}
\end{equation}
with ${A=\dfrac{\lan\kappa^{\alpha}_i|n_i=1\ran}{\lan\kappa_i|n_i=1\ran}}$ and ${B=A\lan\kappa_i|X_i=1\ran-\lan\kappa^\alpha_i|X_i=1\ran}$. Note that~$A>0$.

\subsection{Mean-field approximation}
Assuming that the state of the nodes are uncorrelated, we can write
\begin{equation}
\text{Pr}(X_i=1)=\text{Pr}(n_i=0\cap\delta_{z_i}^1=1)\approx\text{Pr}(n_i=0)\times\text{Pr}(\delta_{z_i}^1=1)\ ,
\end{equation}
with ${\text{Pr}(n_i=0)=1-\lan n_i\ran}$ and
\begin{equation}
\text{Pr}(\delta_{z_i}^1=1)\approx{k_i\choose1}\times\text{Pr}(n_i=1)\times[\text{Pr}(n_i=0)]^{k_i-1}=k_i\lan n_i\ran\left(1-\lan n_i\ran\right)^{k_i-1}\ .
\end{equation}
Additionally, for uncorrelated networks we can write the topology term as
\begin{equation}
\lan(1-n_i)z_i\ran=\lan(1-n_i)\sum_ja_{ij}n_j\ran=\sum_ja_{ij}(\lan n_j\ran-\lan n_in_j\ran)\approx\sum_j\frac{k_ik_j}{N\lan k\ran}(\lan n_j\ran-\lan n_i\ran\lan n_j\ran)\ .
\end{equation}
In random degree-regular networks we have ${k_i=k_j=\lan k\ran=k}$, and after applying the mean-field approximation, ${\lan n_i\ran=\lan n_j\ran=\rho}$, Eq.~\eqref{eq:singleeq}  becomes
\begin{equation}
-\mu^{-\alpha}\rho+A\upsilon\eta^{-1}k\rho(1-\rho)-Bk\rho(1-\rho)^k=0\ .
\end{equation}

For $\rho\approx0$ it is reasonable to assume that the coefficients $A$ and $B$ are constant. For simplicity we use the reduced coefficients ${a=\mu^\alpha\upsilon\eta^{-1}A}$ and ${b=\mu^\alpha B}$ (see section~\ref{subsec:coefficients0} for details on recasting $a$ and $b$ in terms of the original parameters). The dynamics of the system is then encapsulated by the function
\begin{equation}
f(\rho)=-\rho+ak\rho(1-\rho)-bk\rho(1-\rho)^k\label{eq:ffunc}
\end{equation}
and its first and second derivatives,
\begin{align}
f'(\rho)&=-1+ak(1-2\rho)-bk(1-\rho)^k+bk^2\rho(1-\rho)^{k-1}\\
f''(\rho)&=-2ak+2bk^2(1-\rho)^{k-1}-bk^2(k-1)\rho(1-\rho)^{k-2}\ .
\end{align}
The fixed points, $\rho^*$, are given by ${f(\rho^*)=0}$. Linear stability analysis reveals that these are stable when ${f'(\rho^*)<0}$ and unstable when ${f'(\rho^*)>0}$. Moreover, the transition is continuous if ${f''(\rho^*)<0}$ and discontinuous if ${f''(\rho^*)>0}$.

\subsection{Phase diagram}
From Eq.~\eqref{eq:ffunc} we find that the healthy phase ${\rho^*=0}$ is always a fixed point. It is stable for ${b> a-1/k}$ and unstable for ${b<a-1/k}$. The nature of the transition changes at ${b=a/k}$, and the intersection with ${b=a-1/k}$ yields a tricritical point located at ${a_\text{tc}=1/(k-1)}$, ${b_\text{tc}=1/k(k-1)}$. Then, the phase transition is continuous for ${b<b_\text{tc}}$ and discontinuous for ${b>b_\text{tc}}$. The endemic phase is found by solving
\begin{equation}
-1+ak(1-\rho)-bk(1-\rho)^k=0\ .\label{eq:endemicphase}
\end{equation}
Near the critical point ${b=a_\text{c}-1/k}$, ${\rho\approx0}$ and we can expand ${(1-\rho)^k=1-k\rho+k(k-1)\rho^2+\mathcal{O}(\rho)^3}$. For ${b<b_\text{tc}}$,  Eq.~\eqref{eq:endemicphase} becomes
\begin{equation}
(a-a_\text{c})k-ak\rho+(a_\text{c}k-1)\rho+\mathcal{O}(\rho)^2=0\ ,
\end{equation}
and the prevalence scales as ${\rho\propto(a-a_\text{c})^{\beta_\text{c}}}$ with ${\beta_\text{c}=1}$, the customary mean-field exponent~\cite{Pastor-Satorras:2015qf}. For ${b=b_\text{tc}}$, Eq.~\eqref{eq:endemicphase} is of the form
\begin{equation}
(a-a_\text{c})k-ak(k-1)\rho^2+\mathcal{O}(\rho)^3=0\ ,
\end{equation}
and the prevalence scales as ${\rho\propto(a-a_\text{c})^{\beta_\text{tc}}}$ with ${\beta_\text{tc}=1/2}$.

\subsection{Detailed derivation of dynamic equations}\label{subsec:details0}
To obtain Eq.~\eqref{eq:dkdt0}, the dynamic equation of $\lan\kappa_i\ran(t)$, we start from
\begin{align}
\kappa_i(t+\text{d}t)={ }&\kappa_i(t)-\kappa_i(t)n_i(t)\xi_i+(1-n_i(t))(1-\delta_{z_i(t)}^0)(1-\pi_i)(1-\chi_i)\upsilon z_i(t)\text{d}t\nonumber\\
&-\kappa_i(t)(1-n_i(t))(1-\delta_{z_i(t)}^0)(1-\pi_i)\chi_i\ ,
\end{align}
for which we compute the expectation value conditioned on time~$t$, which only affects the stochastic variables
\begin{align}
\text{E}[\kappa_i(t+\text{d}t)|t]={ }&\kappa_i(t)-\kappa_i(t)n_i(t)\text{E}[\xi_i]+(1-n_i(t))(1-\delta_{z_i(t)}^0)\text{E}[(1-\pi_i)]\text{E}[(1-\chi_i)]\upsilon z_i(t)\text{d}t\nonumber\\
&-\kappa_i(t)(1-n_i(t))(1-\delta_{z_i(t)}^0)\text{E}[(1-\pi_i)]\text{E}[\chi_i]\\
={ }&\kappa_i(t)-\kappa_i(t)n_i(t)\eta\dt+(1-n_i(t))(1-\delta_{z_i(t)}^0)(1-\omega_i(t)\dt)(1-\delta_{z_i(t)}^1\eta\dt)\upsilon z_i(t)\dt\nonumber\\
&-\kappa_i(t)(1-n_i(t))(1-\delta_{z_i(t)}^0)(1-\omega_i(t)\dt)\delta_{z_i(t)}^1\eta\dt\\
={ }&\kappa_i(t)-\kappa_i(t)n_i(t)\eta\dt+(1-n_i(t))(1-\delta_{z_i(t)}^0)\upsilon z_i(t)\dt\nonumber\\
&-\kappa_i(t)(1-n_i(t))(1-\delta_{z_i(t)}^0)\delta_{z_i(t)}^1 \eta\dt+\mathcal{O}(\dt)^2\\
={ }&\kappa_i(t)-\kappa_i(t)n_i(t)\eta\dt+(1-n_i(t))\upsilon z_i(t)\dt-\kappa_i(t)(1-n_i(t))\delta_{z_i(t)}^1\eta\dt+\mathcal{O}(\dt)^2\ .
\end{align}
Next we take the ensemble average up to~$\mathcal{O}(\dt)$
\begin{equation}
\lan\kappa_i(t+\dt)\ran=\lan\text{E}[\kappa_i(t+\text{d}t)|t]\ran=\lan\kappa_i(t)\ran-\eta\lan\kappa_i(t)n_i(t)\ran\dt-\eta\lan\kappa_i(t)(1-n_i(t))\delta_{z_i(t)}^1\ran\dt+\upsilon\lan(1-n_i(t))z_i(t)\ran\dt\ ,
\end{equation}
from where we find
\begin{equation}
\frac{\text{d}\lan\kappa_i(t)\ran}{\dt}=\frac{\langle\kappa_i(t+\dt)\rangle-\langle\kappa_i(t)\ran}{\dt}=-\eta\lan\kappa_i(t)n_i(t)\ran-\eta\lan\kappa_i(t)(1-n_i(t))\delta_{z_i(t)}^1\ran+\upsilon\lan(1-n_i(t))z_i(t)\ran\ .
\end{equation}

In order to obtain Eq.~\eqref{eq:dkgdt0}, the dynamic equation of $\lan\kappa^\gamma_i\ran(t)$, we start from
\begin{align}
\kappa^\gamma_i(t+\text{d}t)={ }&\left[\kappa_i(t)-\kappa_i(t)n_i(t)\xi_i+(1-n_i(t))(1-\delta_{z_i(t)}^0)(1-\pi_i)(1-\chi_i)\upsilon z_i(t)\text{d}t\right.\nonumber\\
&\left.-\kappa_i(t)(1-n_i(t))(1-\delta_{z_i(t)}^0)(1-\pi_i)\chi_i\right]^\gamma\ .
\end{align}
In this case, the computation of the expectation value conditioned on time~$t$ is a little more involved, and is given by the generic expression
\begin{align}
\text{E}[\kappa_i^\gamma(t+\dt)|t]={ }&\text{Pr}(\xi_i=0,\pi_i=0,\chi_i=0)\times \kappa^\gamma_i(t+\dt;\xi_i=0,\pi_i=0,\chi_i=0)\nonumber\\
&+\text{Pr}(\xi_i=0,\pi_i=0,\chi_i=1)\times \kappa^\gamma_i(t+\dt;\xi_i=0,\pi_i=0,\chi_i=1)\nonumber\\
&+\text{Pr}(\xi_i=0,\pi_i=1,\chi_i=0)\times \kappa^\gamma_i(t+\dt;\xi_i=0,\pi_i=1,\chi_i=0)\nonumber\\
&+\text{Pr}(\xi_i=1,\pi_i=0,\chi_i=0)\times \kappa^\gamma_i(t+\dt;\xi_i=1,\pi_i=0,\chi_i=0)\nonumber\\
&+\mathcal{O}(\dt)^2\label{eq:kggeneric}\ .
\end{align}
Computing the necessary terms
\begin{align}
\text{Pr}(\xi_i=0,\pi_i=0,\chi_i=0)&=1-(\eta+\omega_i(t)+\delta_{z_i(t)}^1\eta)\dt+\mathcal{O}(\dt)^2\\
\text{Pr}(\xi_i=0,\pi_i=0,\chi_i=1)&=\delta_{z_i(t)}^1\eta\dt+\mathcal{O}(\dt)^2\\
\text{Pr}(\xi_i=0,\pi_i=1,\chi_i=0)&=\omega_i(t)\dt+\mathcal{O}(\dt)^2\\
\text{Pr}(\xi_i=1,\pi_i=0,\chi_i=0)&=\eta\dt+\mathcal{O}(\dt)^2
\end{align}
\begin{align}
\kappa^\gamma_i(t+\dt;\xi_i=0,\pi_i=0,\chi_i=0)&=[\kappa_i(t)+(1-n_i(t))(1-\delta_{z_i(t)}^0)\upsilon z_i(t)\dt]^\gamma\\
&=[\kappa_i(t)+(1-n_i(t))\upsilon z_i(t)\dt]^\gamma\\
&=\kappa^\gamma_i(t)\lbr1+\frac{(1-n_i(t))\upsilon z_i(t)\dt}{\kappa_i(t)}\rbr^\gamma\\
&=\kappa^\gamma_i(t)\lbr1+\frac{\gamma(1-n_i(t))\upsilon z_i(t)\dt}{\kappa_i(t)}\rbr+\mathcal{O}(\dt)^2\\
\kappa^\gamma_i(t+\dt;\xi_i=0,\pi_i=0,\chi_i=1)&=\lbr\kappa_i(t)-\kappa_i(t)(1-n_i(t))(1-\delta_{z_i(t)}^0)\rbr^\gamma\nonumber\\
&=\kappa_i^\gamma(t)\lbr1-(1-n_i(t))(1-\delta_{z_i(t)}^0)\rbr^\gamma\nonumber\\
&=\kappa_i^\gamma(t)\lbr1-(1-n_i(t))(1-\delta_{z_i(t)}^0)\rbr\\
\kappa^\gamma_i(t+\dt;\xi_i=0,\pi_i=1,\chi_i=0)&=\kappa^\gamma_i(t)\\
\kappa^\gamma_i(t+\dt;\xi_i=1,\pi_i=0,\chi_i=0)&=[\kappa_i(t)-\kappa_i(t)n_i(t)+(1-n_i(t))(1-\delta_{z_i(t)}^0)\upsilon z_i(t)\dt]^\gamma\\
&=[\kappa_i(t)(1-n_i(t))+(1-n_i(t))\upsilon z_i(t)\dt]^\gamma\\
&=(1-n_i(t))^\gamma[\kappa_i(t))+\upsilon z_i(t)\dt]^\gamma\\
&=(1-n_i(t))\kappa_i^\gamma(t)\lbr1+\frac{\upsilon z_i(t)\dt}{\kappa_i(t)}\rbr^\gamma\\
&=(1-n_i(t))\kappa_i^\gamma(t)\lbr1+\frac{\beta\upsilon z_i(t)\dt}{\kappa_i(t)}\rbr+\mathcal{O}(\dt)^2\ ,
\end{align}
and substituting in Eq.~\eqref{eq:kggeneric} yields
\begin{align}
E[\kappa_i^\gamma(t+\dt)|t]={ }&\lbr1-(\eta+\omega_i(t)+\delta_{z_i(t)}^1\eta)\dt\rbr\kappa^\gamma_i(t)\lbr1+\frac{\gamma(1-n_i(t))\upsilon z_i(t)\dt}{\kappa_i(t)}\rbr\nonumber\\
&+(\delta_{z_i(t)}^1\eta\dt)\kappa^\gamma_i(t)(1-(1-n_i(t))(1-\delta_{z_i(t)}^0))+(\omega_i(t)\dt)\kappa^\gamma_i(t)\nonumber\\
&+(\eta\dt)(1-n_i(t))\kappa_i^\gamma(t)\lbr1+\frac{\gamma\upsilon z_i(t)\dt}{\kappa_i(t)}\rbr\\
={ }&\kappa^\gamma_i(t)\lbr1-(\eta+\omega_i(t)+\delta_{z_i(t)}^1\eta)\dt+\frac{\gamma(1-n_i(t))\upsilon z_i(t)\dt}{\kappa_i(t)}\rbr+(\delta_{z_i(t)}^1\eta\dt)\kappa^\gamma_i(t)(1-(1-n_i(t)))\nonumber\\
&+(\omega_i(t)\dt)\kappa^\gamma_i(t)\nonumber + \eta\dt(1-n_i(t))\kappa_i^\gamma(t)+\mathcal{O}(\dt)^2\\
={ }&\kappa_i^\gamma(t)-\eta \kappa_i^\gamma(t)n_i(t)\dt-\eta\kappa_i^\gamma(t)(1-n_i(t))\delta_{z_i(t)}^1\dt+\gamma\upsilon(1-n_i(t))z_i(t)\kappa_i^{\gamma-1}(t)\dt+\mathcal{O}(\dt)^2\ .
\end{align}
Then we take the ensemble average up to~$\mathcal{O}(\dt)$
\begin{equation}
\lan\kappa^\gamma_i(t+\dt)\ran=\lan\kappa^\gamma_i(t)\ran-\eta\lan\kappa^\gamma_i(t)n_i(t)\ran\dt-\eta\lan\kappa_i^\gamma(t)(1-n_i(t))\delta_{z_i(t)}^1\ran\dt+\gamma\upsilon\lan(1-n_i(t))z_i(t)\kappa_i^{\gamma-1}(t)\ran\dt\ ,
\end{equation}
and finally find
\begin{equation}
\frac{\text{d}\lan\kappa^\gamma_i(t)\ran}{\dt}=\frac{\langle\kappa^\gamma_i(t+\dt)\rangle-\langle\kappa^\gamma_i(t)\ran}{\dt}=-\eta\lan\kappa^\gamma_i(t)n_i(t)\ran-\eta\lan\kappa_i^\gamma(t)(1-n_i(t))\delta_{z_i(t)}^1\ran+\gamma\upsilon\lan(1-n_i(t))z_i(t)\kappa^{\gamma-1}_i(t)\ran\ .
\end{equation}

\subsection{Recasting into original parameters}\label{subsec:coefficients0}
In order to obtain the coefficient ${A=\lan\kappa^{\alpha}_i|n_i=1\ran/\lan\kappa_i|n_i=1\ran}$, we need to compute ${\lan\kappa^{\gamma}_i|n_i=1\ran}$ for ${\gamma=1}$ and ${\gamma=\alpha}$. These moments are conditioned on ${n_i=1}$, i.e., node~$i$ being infected. Since the viral load does not change while infected,  measuring $\kappa_i$ of infected node~$i$ yields the same results as measuring $\kappa_i$ at the moment~$i$ became infected. Within the mean-field approximation, these quantities are the same for all nodes, hence we drop the $i$~index. We denote the required density by $\phi(\kappa)$, i.e., the probability that a node had amassed $\kappa$ viral load at the moment it became infected.

The difference between the infection probability, $\psi^*(\kappa)$, and $\phi(\kappa)$ is subtle but crucial in the case of short-term memory. Recall that with $\zeta=0$, a susceptible node instantly erases its viral load when it becomes dormant. Thus, reaching a viral load of $\kappa$ is conditioned on being continuously exposed to the pathogen. Simply put, $\psi^*(\kappa)$ measures the probability of becoming infected when the accumulated viral load is~$\kappa$, while $\phi(\kappa)$ measures the probability of reaching an accumulated viral load of $\kappa$ and then becoming infected.

When $\rho\approx0$ we can assume that susceptible node~$i$ has only one infected neighbor~$j$. Moreover, $i$ transitioned from dormant to susceptible at the same time that $j$ became infected. Thus, node~$i$ is exposed to a single source of pathogen, and the time since $j$ became infected, $t$, and the viral load accumulated by~$i$, $\kappa$, are proportional, ${\kappa=\upsilon t}$. Then $\phi(\kappa)$ is the probability of $i$ becoming infected with~$\kappa$, $\psi_\text{inf}^*(\kappa)$, times the probability that $j$ does not recover before~$t$, $\Psi_\text{rec}(t)$. Expressed in terms of $\kappa$ this is
\begin{equation}
\phi(\kappa)=N^{-1}\psi_\text{inf}(\kappa)\Psi_\text{rec}(\kappa)=N^{-1}\alpha\mu^\alpha\kappa^{\alpha-1}e^{-(\mu\kappa)^\alpha}e^{-\eta\upsilon^{-1}\kappa}\ ,
\end{equation}
with ${N=\int_0^\infty\psi_\text{inf}(\kappa)\Psi_\text{rec}(\kappa)\text{d}\kappa}$. Defining the integral
\begin{equation}
I(\theta)=\int_0^\infty\kappa^\theta e^{-(\mu\kappa)^\alpha}e^{-\eta\upsilon^{-1}\kappa}\text{d}\kappa\ ,
\end{equation}
the corresponding moments are
\begin{equation}
\lan\kappa^\gamma|n=1\ran=\frac{I(\gamma+\alpha-1)}{I(\alpha-1)}\ .
\end{equation}

We proceed in a similar manner for the computation of the moments ${\lan\kappa^\gamma_i|X_i=1\ran}$. Now we are sampling the state of a node that is susceptible and has only one infected neighbor. For $\rho\approx0$ we assume the same scenario as before: since becoming susceptible, node~$i$ has been exposed continuously to a single infected neighbor~$j$. Node~$i$'s state becomes $X_i=0$ either when $j$ recovers or when $i$ becomes infected itself. This event occurs when $i$ has accumulated $\kappa$ viral load with probability density
\begin{equation}
\varphi(\kappa)=N^{-1}\lbr\psi_\text{inf}(\kappa)\Psi_\text{rec}(\kappa)+\Psi_\text{inf}(\kappa)\psi_\text{rec}(\kappa)\rbr=N^{-1}\lbr\alpha\mu^\alpha\kappa^{\alpha-1}+\eta\upsilon^{-1}\rbr e^{-(\mu\kappa)^\alpha}e^{-\eta\upsilon^{-1}\kappa}\ ,
\end{equation}
with ${N=\int_0^\infty\lbr\psi_\text{inf}(\kappa)\Psi_\text{rec}(\kappa)+\Psi_\text{inf}(\kappa)\psi_\text{rec}(\kappa)\rbr\text{d}\kappa}$. Then the probability density to sample node~$i$ in state ${X_i=1}$ with a viral load of $\kappa$ is~\cite{cox:1970}
\begin{equation}
\phi(\kappa)=\frac{\Phi(\kappa)}{\lan\kappa\ran_{\varphi}}\ ,
\end{equation}
with 
\begin{align}
\Phi(\kappa)&=N^{-1}\int_{\kappa}^\infty\lbr\alpha\mu^\alpha s^{\alpha-1}+\eta\upsilon^{-1}\rbr e^{-(\mu s)^\alpha}e^{-\eta\upsilon^{-1}s}\text{d}s\\
\lan\kappa\ran_{\varphi}&=N^{-1}\int_{0}^\infty\lbr\alpha\mu^\alpha\kappa^{\alpha}+\eta\upsilon^{-1}\kappa\rbr e^{-(\mu\kappa)^\alpha}e^{-\eta\upsilon^{-1}\kappa}\text{d}\kappa\ .
\end{align}
Defining the integrals
\begin{align}
J_1(\kappa)&=\int_{\kappa}^\infty\lbr\alpha\mu^\alpha s^{\alpha-1}+\eta\upsilon^{-1}\rbr e^{-(\mu s)^\alpha}e^{-\eta\upsilon^{-1}s}\text{d}s\\
J_2&=\int_{0}^\infty\lbr\alpha\mu^\alpha\kappa^{\alpha}+\eta\upsilon^{-1}\kappa\rbr e^{-(\mu\kappa)^\alpha}e^{-\eta\upsilon^{-1}\kappa}\text{d}\kappa\ ,
\end{align}
the corresponding moments are
\begin{equation}
\lan\kappa^\gamma|X=1\ran=\frac{1}{J_2}\int_0^\infty\kappa^\gamma J_1(\kappa)\text{d}\kappa\ .
\end{equation}

Figure 3 of the main text is obtained as follows. For a given $\lambda$ and~$\alpha$, we compute the moments ${\lan\kappa^\gamma|n=1\ran}$ and ${\lan\kappa^\gamma|X=1\ran}$ with $\eta=\upsilon=1$. The integrals are compute numerically using the \texttt{Python} package \texttt{SciPy}. With these results we compute the coefficients $A$ and~$B$, and from there the coefficients $a$ and~$b$. With~$k=4$, the critical point $\lambda_\text{c}$ is computed finding the root of ${-1+ak-bk=0}$. The sign of ${bk^2-ak}$ marks the nature of the transition. 

\section{Analytics: Long-term memory}
With $\zeta=\infty$, susceptible node~$i$ that becomes dormant freezes its viral load, i.e., it does not reset $\kappa_i$ to zero. Then the equation for ${\kappa_i(t+\dt)}$ reduces to
\begin{equation}
\kappa_i(t+\text{d}t)=\kappa_i(t)-\kappa_i(t)n_i(t)\xi_i+(1-n_i(t))(1-\delta_{z_i(t)}^0)(1-\pi_i)(1-\chi_i)\upsilon z_i(t)\text{d}t\ ,
\end{equation}
where the first term corresponds to the previously amassed viral load, the second term describes the event where infected node~$i$ recovers and erases its viral load, and the third term corresponds to susceptible node~$i$ remaining susceptible (neither recovering nor becoming dormant) and accumulating additional viral load from its $z_i(t)$ infected neighbors (from each at rate~$\upsilon$). Following the same procedure as in the previous section, we find the dynamic equation
\begin{equation}
\frac{\text{d}\lan\kappa_i(t)\ran}{\dt}=-\eta\lan\kappa_i(t)n_i(t)\ran+\upsilon\lan(1-n_i(t))z_i(t)\ran\label{eq:dkdt8}\ ,
\end{equation}
and taking the late-time limit yields
\begin{equation}
0=-\eta\lan\kappa_in_i\ran+\upsilon\lan(1-n_i)z_i\ran\ .\label{eq:ssk8}
\end{equation}

As before, we expand ${\lan\kappa_in_i\ran=\lan\kappa_i|n_i=1\ran\lan n_i\ran}$. The conditioned average ${\lan\kappa_i|n_i=1\ran}$ measures the viral load of an infected node, which again equates to the viral load at infection. With $\zeta=\infty$, nothing hinders the accumulation of viral load, i.e., nodes may freely accumulate any value of~$\kappa$. Thus, the probability of having $\kappa$ at the moment of infection is simply the probability of becoming infected with~$\kappa$, i.e., $\psi^*(\kappa)$. Using this probability density, ${\lan\kappa_i|n_i=1\ran=\lan\kappa\ran_\text{inf}}$, the value used in the definition of the effective spreading ratio, ${\lambda=\upsilon\lan t\ran_\text{rec}/\lan\kappa\ran_\text{inf}}$. Substituting in Eq.~\eqref{eq:ssk8} gives
\begin{equation}
\lan n_i\ran=\lambda\lan(1-n_i)z_i\ran\ .
\end{equation}
Note that this equation is identical to the result obtained for the standard SIS~\cite{Pastor-Satorras:2015qf}, which demonstrates that the prevalence is independent of~$\alpha$, and equivalent to the Markovian model.

\subsection{Additional verification}
The value of $\zeta$ does not directly affect the state of node~$i$, thus the dynamic equation for $n_i(t)$ is the same as before
\begin{equation}
\frac{\text{d}\lan n_i(t)\ran}{\dt}=-\eta\lan n_i(t)\ran+\upsilon\alpha\mu^\alpha\lan(1-n_i(t))z_i(t)\kappa_i^{\alpha-1}(t) \ran\ .\label{eq:dndt8}
\end{equation}
In order to obtain the dynamic equation for $\lan\kappa_i^\gamma(t)\ran$ we start from 
\begin{equation}
\kappa^\gamma_i(t+\text{d}t)=\left[\kappa_i(t)-\kappa_i(t)n_i(t)\xi_i+(1-n_i(t))(1-\delta_{z_i(t)}^0)(1-\pi_i)(1-\chi_i)\upsilon z_i(t)\text{d}t\right]^\gamma\ ,
\end{equation}
and employ Eq.~\eqref{eq:kggeneric} to compute the expectation value conditioned on time~$t$. All the terms are identical as before, except for
\begin{equation}
\kappa^\gamma_i(t+\dt;\xi_i=0,\pi_i=0,\chi_i=1)=\kappa^\gamma_i(t)\ ,
\end{equation}
then
\begin{equation}
E[\kappa_i^\gamma(t+\dt)|t]=\kappa_i^\gamma(t)-\eta \kappa_i^\gamma(t)n_i(t)\dt+\gamma\upsilon(1-n_i(t))z_i(t)\kappa_i^{\beta-1}(t)\dt\ .
\end{equation}
Finally, computing the ensemble average and rearranging the equation yields
\begin{equation}
\frac{\text{d}\lan\kappa^\gamma_i(t)\ran}{\dt}=-\eta\lan\kappa^\gamma_i(t)n_i(t)\ran +\gamma\upsilon\lan(1-n_i(t))z_i(t)\kappa^{\gamma-1}_i(t)\ran\ .\label{eq:dkgdt8}
\end{equation}

Setting $\gamma=\alpha$ in Eq.~\eqref{eq:dkgdt8}, taking the late-time limit in Eqs.~\eqref{eq:dndt8} and~\eqref{eq:dkgdt8}, and dropping the dependence with $t$ gives
\begin{align}
0&=-\eta\lan n_i\ran+\upsilon\alpha\mu^\alpha\lan(1-n_i)z_i\kappa^{\alpha-1}_i\ran\\
0&=-\eta\lan\kappa_i^\alpha n_i\ran+\alpha\upsilon\lan(1-n_i)z_i\kappa_i^{\alpha-1}\ran\ .
\end{align}
Using the expansion ${\lan\kappa_i^\alpha n_i\ran=\lan\kappa_i^\alpha|n_i=1\ran\lan n_i\ran}$, and combining both equations yields ${\lan\kappa_i^\alpha|n_i=1\ran=\mu^{-\alpha}}$, which recovers the result of
\begin{equation}
\lan\kappa^\alpha\ran_\text{inf}=\int_0^\infty\kappa^\alpha\psi^*(\kappa)\text{d}\kappa=\int_0^\infty\alpha\mu^\alpha\kappa^{2\alpha-1}e^{-(\mu\kappa)^\alpha}\text{d}\kappa=\mu^{-\alpha}\ ,
\end{equation}
and validates the computation of $\lan\kappa_i|n_i=1\ran$ using $\psi^*(\kappa)$ .

\section{Simulations}
\subsection{Core algorithm}
Here we provide a schematic outline of the core simulation algorithm, for fixed values of $\alpha$, $\mu$, $\upsilon$, $\eta$, $\lambda$, and~$\zeta=0$ or $\zeta=\infty$. At a given time~$t$, the nodes are separated in three lists: infected (I), dormant (D), and susceptible (S). For dormants we store their accumulated viral load~$\kappa$. For susceptibles we store $\kappa$ and also the number of infected neighbors~$z$. The contact network is encoded in an adjacency matrix or list.
\begin{enumerate}
\item Sample the interval $\tau$, solving ${\Phi(\tau)=u}$, with $u\in U(0,1)$ and
\begin{equation}
\Phi(\tau)=\prod_{i\in\text{I}} e^{-\eta\tau}\prod_{j\in\text{S}}\frac{e^{-[\mu(\kappa_j+\upsilon z_j\tau)]^\alpha}}{e^{-(\mu\kappa_j)^\alpha}}
\end{equation}
\item Update the viral load of susceptible nodes, ${\kappa_j\leftarrow\kappa_j+z_j\tau}$.
\item Compute the hazard rate for infected nodes, $\omega_i=\eta$, the hazard rate for susceptible nodes, ${\omega_j=\upsilon\alpha\mu^\alpha z_j(\kappa_j)^{\alpha-1}}$, and the total hazard rate, ${\Omega=\sum_{i\in\text{I}}\omega_i+\sum_{j\in\text{S}}\omega_j}$.
\item Compute the discrete distribution ${\Pi_k=\omega_k/\Omega}$, and sample the next-occurring event.
\begin{itemize}
\item Infected node~$k$ recovers.
\begin{enumerate}
\item Compute its number of infected neighors, $z_k$.
\begin{itemize}
\item If $z_k=0$, move node~$k$ to the dormant list with $\kappa_k=0$.
\item If $z_k>0$, move node~$k$ to the susceptible list with $\kappa_k=0$ and store $z_k$.
\end{itemize}
\item For all of $k$'s neighbors, decrease the number of infected neighbors by one, ${z_\ell\leftarrow z_\ell-1}$.
\begin{itemize}
\item If $z_\ell=0$, move node~$\ell$ from the susceptible list to the dormant list.
\begin{itemize}
\item If $\zeta=0$, set $\kappa_\ell=0$.
\end{itemize}
\end{itemize}
\end{enumerate}
\item Susceptible node~$k$ becomes infected.
\begin{enumerate}
\item Move node~$k$ to the infected list.
\item For all of $k$'s neighbors, increase the number of infected neighbors by one, ${z_\ell\leftarrow z_\ell+1}$.
\begin{itemize}
\item If $z_\ell=1$, move node~$\ell$ from the dormant list to the susceptible list and store $z_\ell$.
\end{itemize}
\end{enumerate}
\end{itemize}
\end{enumerate}

\subsection{Late-time prevalence curve}
Given the stochastic nature of the dynamics we must average over independent realizations in order to obtain a representative ${\rho_\text{st}(\lambda)}$ curve. Additionally, we want to control the spacing in the $\rho_\text{st}$ axis and handle the increasing correlation time as we approach the critical point. We employ a two-step simulation scheme: in the first, preparatory step we elaborate a list of $\lambda$ values that will be used in the second, sampling step to extensively compute~$\rho_\text{st}$.

\begin{figure}
\includegraphics[width=0.4\textwidth]{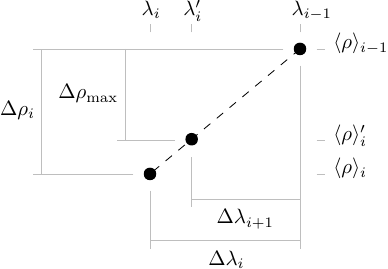}
\caption{Interpolation construction if ${\Delta\rho_i>\Delta\rho_\text{max}}$, yielding $\lambda'_i$, $\langle\rho\rangle'_i$, and $\Delta\lambda_{i+1}$.}
\label{fig:adjustslo}
\end{figure}

For a given network of size~$N$, we start at ${\lambda=\lambda_0}$ and infect all nodes. We evolve the system during ${25\times M_0}$ events (with ${M_0=N}$), record the final value of ${\rho=N_\text{I}/N}$ and store the system's final state. We repeat this for $R$ independent runs, starting each time at $\lambda_0$ and storing each final state separately. We compute the average ${\langle\rho\rangle_0}$ (of the $R$ measures) and write $\lambda_0$ and $M_0$ to the output file. Next we decrease the control parameter, ${\lambda_i=\lambda_{i-1}-\lambda_i}$. For each run we load the corresponding initial state from storage and iterate ${25\times M_i}$ events. Then we record the final value of $\rho$ and store the system's final state. After repeating this for the $R$ runs, we compute $\langle\rho\rangle_i$ and ${\Delta\rho_i=\langle\rho\rangle_{i-1}-\langle\rho\rangle_i}$. If ${\Delta\rho_i>\Delta\rho_\text{max}}$, we interpolate the results (see Fig. \ref{fig:adjustslo}), setting ${\Delta\lambda_{i+1}=\lambda_i\Delta\rho_\text{max}/\Delta\rho_i}$, and reassigning ${\lambda_i\leftarrow\lambda_i+\Delta\lambda_i-\Delta\lambda_{i+1}}$ and ${\langle\rho\rangle_i\leftarrow\langle\rho\rangle_{i-1}-\Delta\rho_\text{max}}$. In addition, we increase the event interval, ${M_{i+1}=M_i^\epsilon}$, with ${\epsilon>1}$. Finally we write $\lambda_i$ and $M_i$ to the output file (note that ${\lambda_i=\lambda_0}$ and ${M_1=M_0}$). A run is deactivated when it reaches the absorbing state (${N_\text{I}=0}$): we stop simulating its dynamics and it is no longer included in the computation of~$\langle\rho\rangle$. We keep decreasing the control parameter $\lambda$ until all runs are deactivated, the point at which the simulation is halted.

Following this preparatory step we proceed with an extensive sampling of the control parameter. We infect all nodes and load $\lambda$ and $M$ from the input file. We thermalize the system during ${20\times M}$ events and afterwards measure $\rho$ and $\rho^2$ of $X$ states, each separated by a window of $M$ events. We repeat this thermalization-sampling procedure for the next entry on the input list, using the last visited state as initial condition. The simulation is halted when the list is fully iterated or whenever the system reaches the absorbing state. We repeat the whole procedure for $Y$ independent runs, each using the same input file. The results are temporally averaged, i.e., each measure $j$ is weighted by its residency time~$\tau_j$: ${\langle\rho\rangle=w^{-1}\sum_{j=1}^{Z}\tau_j\rho_j}$ and ${\langle\rho^2\rangle=w^{-1}\sum_{j=1}^Z\tau_j\rho^2_j}$, with ${w=\sum_{j=1}^Z\tau_j}$ and $Z$ the total number of samples (note that, since the input file may not be fully iterated, ${Z\leq X\times Y}$). We estimate the order parameter as ${\rho_\text{st}=\langle\rho\rangle}$ and compute its standard error as ${s(\rho_\text{st})=\sqrt{(\langle\rho^2\rangle-\langle\rho\rangle^2)/Z}}$. We use ${\lambda_0=1.2}$, ${\Delta\lambda_0=0.05}$, ${\Delta\rho_\text{max}=0.025}$, ${\epsilon=1.01}$, ${X=500}$, and ${Y=20}$. We use ${R=40}$ and ${R=20}$ for networks of ${N=10^3}$ and ${N=10^4}$ nodes, respectively (figures 4 and 6 of main text).

\subsection{Single-seed simulations}
For a given network of size $N$ and a fixed value of $\lambda$ we simulate $Z$ independent runs. Each outbreak starts with a single, randomly chosen infected node. All other nodes are healthy, with zero viral load, and the coverage $K$ is set to zero. During the evolution of the outbreak we keep track of all the nodes' first infection label. Whenever a node is infected for the first time, we change its label and increase the coverage in one unit, ${K\leftarrow K+1}$. An outbreak is terminated when it reaches the absorbing state (finite realization) or when the coverage reaches the threshold, ${K_\text{th}=c_\text{th}N}$ (endemic realization). We record the final values $K$ and $K^2$ of all outbreaks, and count the number of endemic realizations, ${z^{(1)}_\text{end}}$. Afterwards we compute the averages ${\langle K\rangle=Z^{-1}\sum_{j=1}^ZK_j}$ and ${\langle K^2\rangle=Z^{-1}\sum_{j=1}^ZK_j^2}$. We estimate the avarege coverage fraction as ${\bar{c}=\langle K\rangle/N}$ and the endemic probability as ${P_1=z^{(1)}_\text{end}/Z}$, with standard errors ${s(\bar{c})=N^{-1}\sqrt{(\langle K^2\rangle-\langle K\rangle^2)/Z}}$ and ${s(P_1)=\sqrt{P_1(1-P_1)/Z}}$.

To compute $P_3$ we proceed from the same initial setting. Now, however, when an outbreak reaches the coverage threshold we reset the coverage to zero and erase all first infection labels (regardless of wether the node is healthy or infected at that moment). Then we continue evolving the same outbreak, keeping track again of each of the nodes' ``first infection'', changing the labels and increasing the coverage accordingly. When the coverage threshold is reached for a second time, this reset is performed once again. We count the number of realizations that are able to reach the coverage a third time, $z_\text{end}^{(3)}$, and compute ${P_3=z_\text{end}^{(3)}/Z}$ and ${s(P_3)=\sqrt{P_3(1-P_3)/Z}}$. We use ${c_\text{th}=0.75}$ and ${Z=10^4}$ (figures 5, 7 and 8 of main text).

\subsubsection{Late-time prevalence}
To measure the late-time prevalence of endemic outbreaks, $\rho_\text{st}^*$, we only consider realizations that, following the procedure explained above, are able to reach the coverage threshold a total of $m$ times. At this point the outbreak has had sufficient time to evolve towards its active steady-state. Afterwards we measure $\rho$ and $\rho^2$ of $X$ states, each separated by a window of $N$ events. For a given $\lambda$ we record a maximum of $W$ states, running a maximum of $Y$ outbreaks. The results are temporally averaged, i.e., each measure $j$ is weighted by its residency time~$\tau_j$: ${\langle\rho\rangle=w^{-1}\sum_{j=1}^{Z}\tau_j\rho_j}$ and ${\langle\rho^2\rangle=w^{-1}\sum_{j=1}^Z\tau_j\rho^2_j}$, with ${w=\sum_{j=1}^Z\tau_j}$ and $Z$ the total number of samples (note that ${Z\leq W}$). We estimate the late-time prevalence as ${\rho^*_\text{st}=\langle\rho\rangle}$ and compute its standard error as ${s(\rho^*_\text{st})=\sqrt{(\langle\rho^2\rangle-\langle\rho\rangle^2)/Z}}$. We use ${m=10}$, ${X=10^2}$, ${W=10^4}$, and ${Y=10^4}$ (figures 5 and 7 of main text).

\subsubsection{Temporal profile}
To represent the evolution of single-seed outbreaks that reach the coverage threshold, $\bar{\rho}(t)$, we start from the usual initial setting, and store the prevalence $\rho(t)$ at given times, with time interval~$\Delta t$. If the system becomes trapped in the absorbing state before reaching the threshold, we discard the trajectory and start again. If the system is able to surpass the coverage threshold, we continue tracking its evolution while the outbreaks remains active, up to~$t_\text{max}$. For a given network of size $N$ and a fixed value of $\lambda$ we average $Z$ trajectories, at fixed values of~$t$: ${\langle\rho(t)\rangle=Z^{-1}\sum_{j=1}^Z\rho_j(t)}$ and ${\langle\rho^2(t)\rangle=Z^{-1}\sum_{j=1}^Z\rho^2_j(t)}$. We estimate the prevalence as ${\bar{\rho}(t)=\langle\rho(t)\rangle}$ and compute its standard error as $s(\bar{\rho}(t))=\sqrt{(\langle\rho^2(t)\rangle-\langle\rho(t)\rangle^2)/Z}$. We use ${\Delta t=0.1}$, ${t_\text{max}=200}$, and ${Z=100}$ (figure 8 of main text).